\documentclass[%
 reprint,
 amsmath,amssymb,
 aps,
]{revtex4-2}
\usepackage{graphicx}
\usepackage{dcolumn}
\usepackage{bm}

\usepackage{xcolor}\usepackage[version=3]{mhchem}

\usepackage[normalem]{ulem}
\begin{document}
\preprint{}

\title{Olivine annealed up to 1500$^\circ$C: changes traced by polarised IR reflectance and magnetization}
\author{
Daniel Smith$^{1,\dag}$, Donatas Narbutis$^{2,\dag}$, Hsin-Hui Huang$^1,\dag$, Philipp Zanon$^{3,4}$, Michael Boschen$^{5}$, Jitraporn Vongsvivut$^6$, Dominique Appadoo$^6$, Soon Hock Ng$^1$, Haoran Mu$^1$, Tomas Katkus$^1$, Nguyen Hoai An Le$^1$, Dan Kapsaskis$^1$, Andy I.R. Herries$^7$, Vijayakumar Anand$^1$, Meguya Ryu$^{8}$, Junko Morikawa$^{8,9,10}$, Saulius Juodkazis$^{1,9,11}$}

\affiliation{$^1$~Optical Sciences Centre and ARC Training Centre in Surface Engineering for Advanced Materials (SEAM), School of Science, Swinburne University of Technology, Hawthorn, Victoria 3122, Australia}
\affiliation{$^2$~Institute of Theoretical Physics and Astronomy, Faculty of Physics, Vilnius University, Saul\.{e}tekio 9, 10222, Vilnius, Lithuania}
\affiliation{$^3$~Space Robotics Group, Space Technology and Industry Institute, Swinburne University of Technology, Hawthorn, Victoria 3122, Australia}
\affiliation{$^4$~Department of Electrical, Robotics and Biomedical Engineering, Swinburne University of Technology, Hawthorn, Victoria 3122, Australia}
\affiliation{$^5$~Surface Engineering for Advanced Materials (SEAM), Swinburne University, Swinburne University of Technology, Hawthorn, Victoria 3122, Australia}
\affiliation{$^6$~
ANSTO‒Australian Synchrotron, 800 Blackburn Road, Clayton, Victoria 3168, Australia}
\affiliation{$^7$~The Australian Archaeomagnetism Laboratory \& Drimolen Palaeoanthropology Field School Dept., Archaeology and History, La Trobe University, Bundoora, 3086, VIC, Australia}
\affiliation{$^8$~School of Materials and Chemical Technology, Institute of Science Tokyo, 2-12-1, Ookayama, Meguro-ku, Tokyo 152-8550, Japan}
\affiliation{$^9$~World Research Hub (WRH), School of Materials and Chemical Technology, Institute of Science Tokyo, 2-12-1, Ookayama, Meguro-ku, Tokyo 152-8550, Japan}
\affiliation{$^{10}$~Research Center for Autonomous Systems Materialogy (ASMat), Institute of Innovative Research, Institute of Science Tokyo,Yokohama 226-8501, Japan}
\affiliation{$^{11}$~Laser Research Center, Physics Faculty, Vilnius University, Saul\.{e}tekio Ave. 10, 10223 Vilnius, Lithuania}
\email{$\dag$ D.S.,D.N.,H-H.H.: equal contributions.}


\date{\today}

\begin{abstract}
Spectral analysis at the infrared (IR) spectral range is introduced with assignment of synthetic red-green-blue (RGB) colours defined by adjustable wavelength and bandwidth ($\lambda,\Delta\lambda$). The RGB bands were selected at the phase-specific absorbance $A$ or reflectance $R$ bands of olivine and related materials, which can be formed via high temperature annealing (HTA) of natural minerals up to $1500~^\circ$C. Natural olivines were collected from quarry at volcanic site in Mortlake, Victoria, Australia and spectrally characterised during IR-THz spectroscopy beamtime experiments at Australian Synchrotron. Phase changes in HTA natural olivines were traced by correlation of optical IR 4-polarisation spectroscopy, X-ray energy dispersive spectroscopy and magnetisation. After HTA, olivine samples were magnetized via precipitation of Fe-rich oxides.    
\end{abstract}

\maketitle

\tableofcontents
\section{Introduction}

Olivine, \ce{(MgFe)2SiO4}, is a hard, green coloured mineral that forms under the 
conditions present in volcanic eruptions~\cite{Wallace-AEPS-2021}, within the oceanic crust and lithosphere, and in the upper mantle. It makes up a major portion 
of the Earth~\cite{Horgan-I-2014}. These magnesium (Mg)--iron (Fe) silicates have important ecological effects, from sequestering \ce{CO2} to hindering climate change~\cite{Schuiling-NS-2017,Hangx-IJGGC-2009}, to reducing ocean acidification~\cite{Schuiling-ESDD-2017}. 

These minerals have also given astronomers insight into the composition of the Universe as olivines are found in asteroids showing that it is a common substance amongst meteorites, and exoplanets alike~\cite{Radomsky-GCA-1990,Pitman-MNRAS-2010}. Since the Earth has 
a large content of olivine ($\sim 60\%$ throughout the upper mantle)~\cite{ringwood1975composition}, it is a good comparison point to other potential planets which could support life. Mars is theorised to have once had the ability to support life, and Mars contains a large percentage of olivine in its crust, much like that of the Earth~\cite{Horgan-I-2014}. Olivine is not stable in the presence of water and changes to iddingsite (the Goldich dissolution series)~\cite{Kuebler}. Finding iddingsite on Mars could help to determine if liquid water once existed on it~\cite{SWINDLE} and is hypothesized to be a way of determining water on other planets. 

The exclusively magnesium silicates have also been discovered to be the main component of interstellar dust, which has large infrared absorbance~\cite{Pitman-MNRAS-2010}, hindering infrared astronomy in regions of high interstellar dust density. 

Olivines come in many different forms based on Fe-to-Mg ratio, from fayalite \ce{Fe2SiO4} with an orthorhombic unit lattice (yellow to brown in colour) to forsterite \ce{Mg2SiO4}, which also has an orthorhombic unit lattice. The wadsleyite (orthorhombic) and ringwoodite (isometric) are the high-pressure polymorphs of olivine. Related minerals make a wider family and include hematite $\alpha-$\ce{Fe2O3}, which has a rhombohedral lattice (brown to reddish-brown), magnetite \ce{Fe3O4} with equal amounts of iron(II) and iron(III), which is known to be ferrimagnetic and has a face-centered cubic lattice, maghemite $\gamma$-\ce{Fe2O3}, which is also ferrimagnetic and has a cubic spinel crystal structure. 
The olivines' colour from light-to-dark green represents how much iron is present in the material, with darker colour showing a larger amount of Fe. 
In its raw form, olivines come as geodes with a hard outer crust that is black and tough to break~\cite{Wallace-AEPS-2021}. This is most likely formed in the eruption process of a volcano with its high temperatures and, when expelled, undergoes rapid cooling to form this compressed outer shell~\cite{Wallace-AEPS-2021}. The inside, still hot, contains high pressure, which causes the glass-like structure to form.

\begin{figure*}[tbh]
\centering\includegraphics[width=18.5cm]{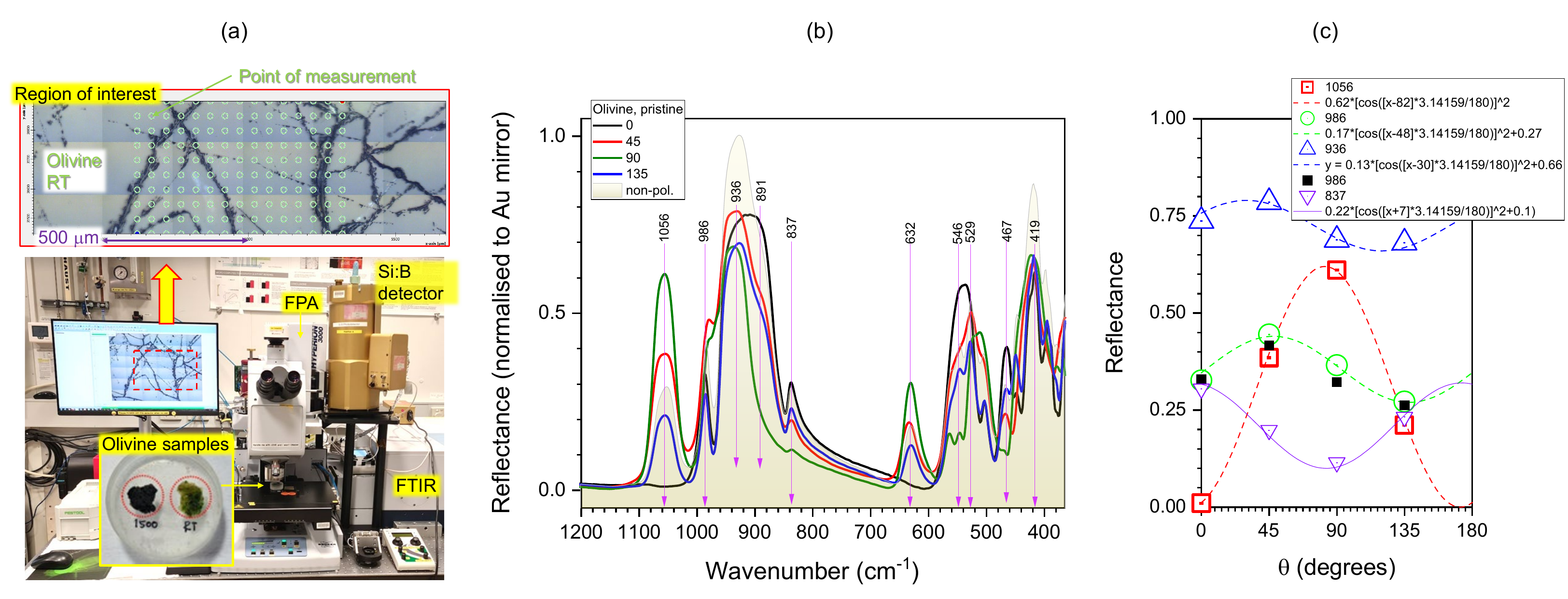}
  \caption{(a) Far-IR microspectroscopic setup at the Australian Synchrotron's IRM beamline using a Si:B photodetector. Region-of-Interest (RoI) was selected and scanned point-by-point in reflection mode. The top inset shows an optical image of the RoI from a polished olivine sample (pristine, RT). (b) IR reflection spectra of all averaged RoI matrix at four polarisations; $0^\circ$ orientation is horizontal (x-axis). Note: several characteristic wavenumbers with reflectance features are marked by down-pointing arrows. The $R\propto\Tilde{\nu}$ dependence is presented in the same way as $R\propto\lambda$ used for automated analysis. (c) Orientation changes at the selected reflectance wavenumbers are fitted by $\mathrm{Amp}\times\cos^2(\theta -\theta_R)+\mathrm{Off}$, where $\theta_R$ is the orientation angle for the maximum reflection, $\mathrm{Amp,Off}$ are the amplitude and offset of the fit; see the exact parameters in the legend. }\label{f-oliv}
\end{figure*}
\begin{figure*}[tbh]
\centering\includegraphics[width=12.5cm]{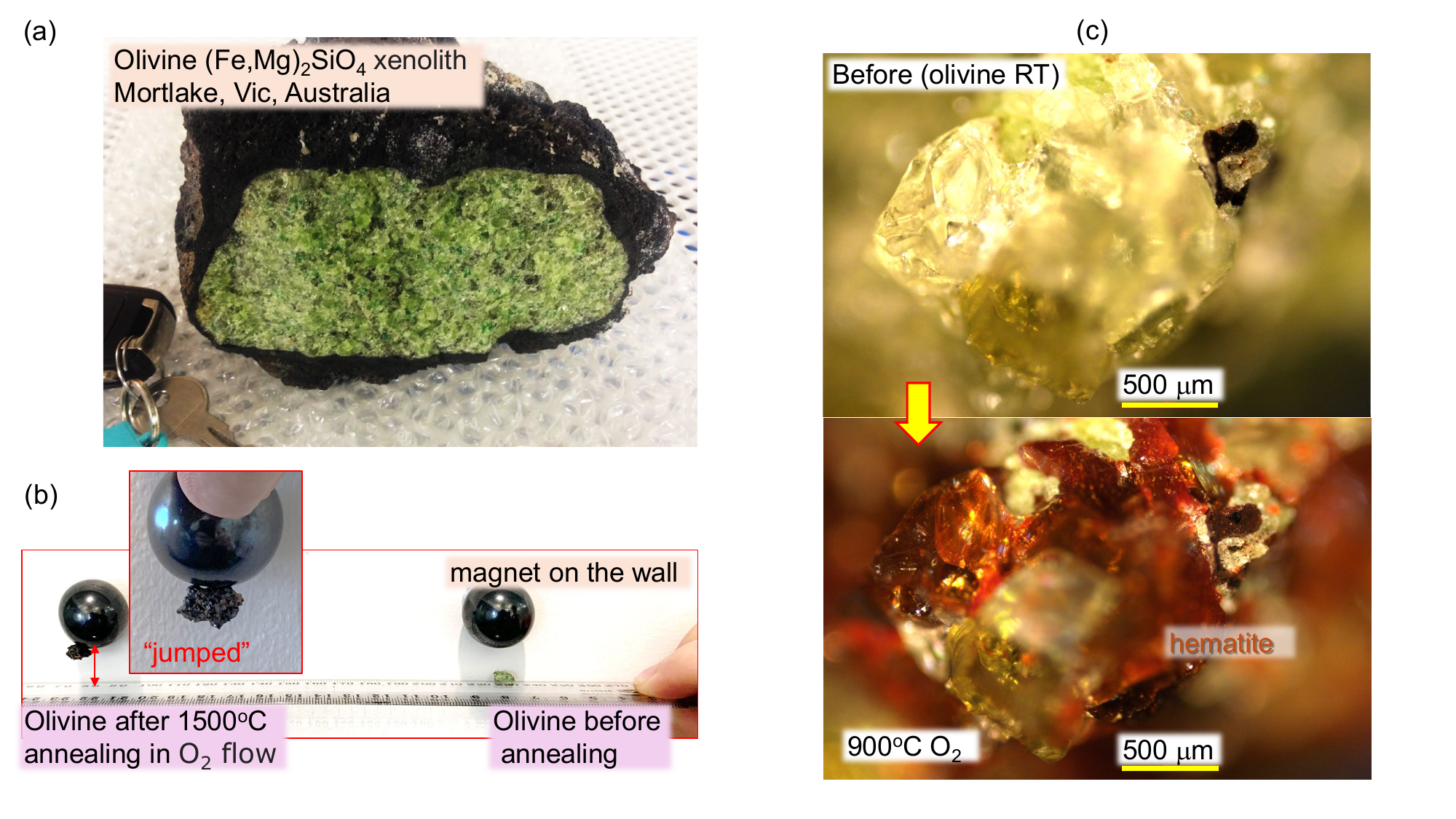}
  \caption{Non-polished samples. (a) Cracked opened geode with olivine interior was used in this study. (b) Olivine turned magnetic after HTA at 1500$^\circ$C for 2~hours in \ce{O2} atmosphere. (c) Change of composition by HTA 900$^\circ$C 2~hours in \ce{O2} atmosphere; same sample, same location. Formation of red-tinted hematite is expected via: maghemite($\gamma$-\ce{Fe2O3}) + 700$^\circ$C = hematite($\alpha$-\ce{Fe2O3}) + magnetite(\ce{Fe3O4})~\cite{multi}. 
    }\label{f-anne}
\end{figure*}
\begin{figure*}[tbh]
\centering\includegraphics[width=18cm]{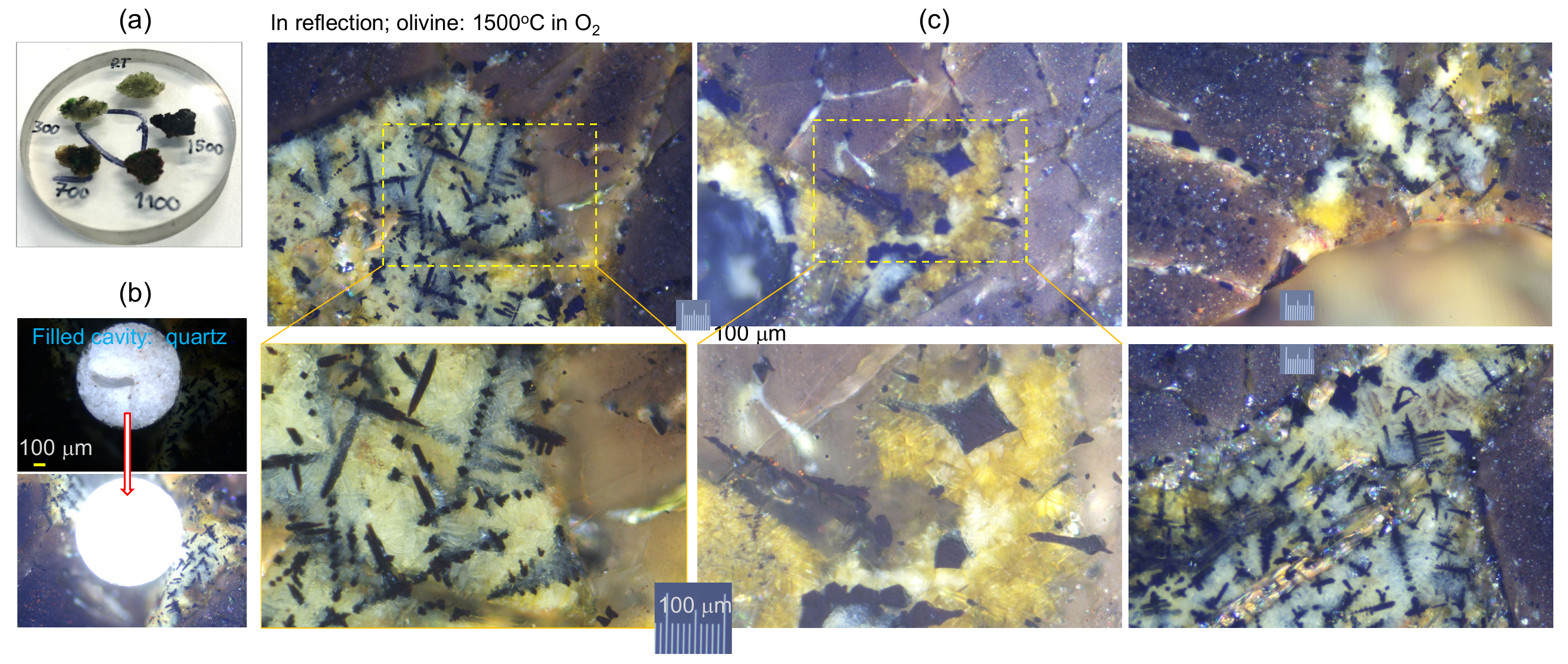}
  \caption{Polished samples. (a) A polished epoxy puck with cut and polished olivines prepared at different HTA conditions for IR imaging in reflection. (b) Quartz micro-cavity in olivine HTA at 1500$^\circ$C for 2~hours in \ce{O2} atmosphere. Two images acquired at different exposures to highlight internal structures. (c) Olivine after HTA 1500$~^\circ$C 2~hours in \ce{O2} atmosphere: different representative regions at different magnification. Dendrite dark regions are representative in samples that showed magnetization; similar dark-red dendritic subsurface structures are present in the HTA samples after 1500$^\circ$C treatment.  
    }\label{f-poli}
\end{figure*}
\begin{figure*}[tbh]
\centering\includegraphics[width=14.5cm]{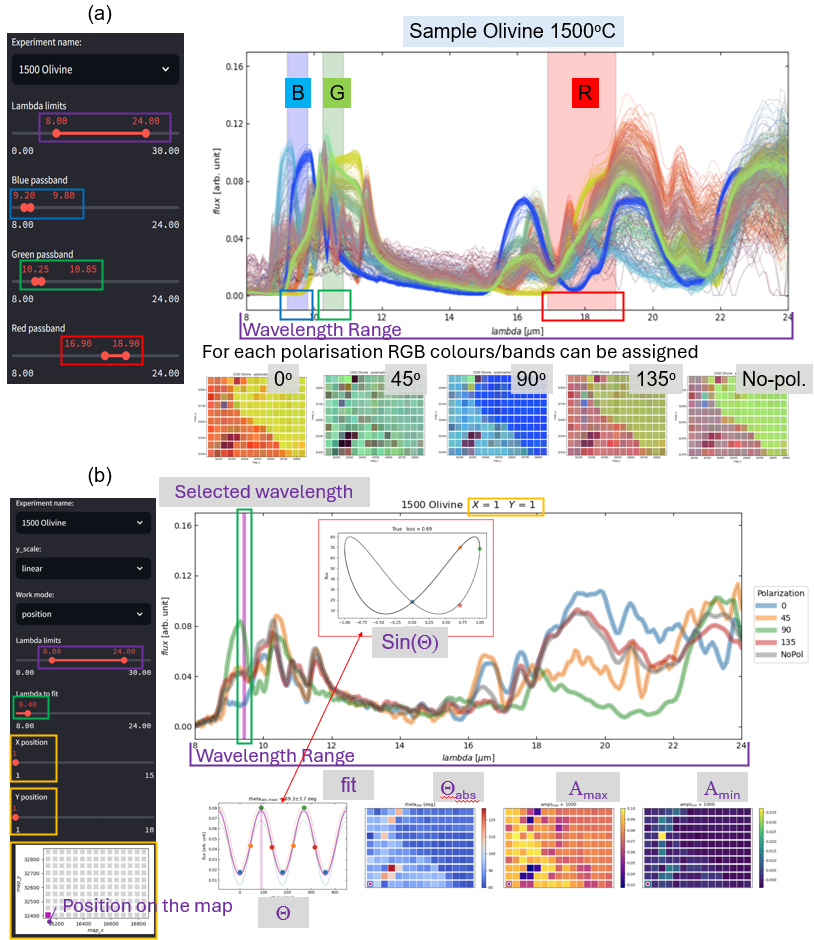}
  \caption{Software for data analysis (Python-based coding). (a) Selection of wavelength $\lambda$~[$\mu$m] bands to assign them an artificial RGB naming (the corresonding wavenumber is $\tilde{\nu} = 1/\lambda$~[cm$^{-1}$]). The 4-polarisation maps (also a non-polarised map) $10\times 15$~pixels can be visualised using an RGB mixture convention. (b) Selected pixel from the map can be analysed for specific absorptance (or reflectance) orientation using fit: $\frac{A_{max-A_{min}}}{2}\cos(2\theta-2\theta_{A,R}) + \frac{A_{max}+A_{min}}{2}$.}\label{f-soft}
\end{figure*}

When olivine samples were heated in a furnace above $\sim 1200^\circ$C and in an oxygen environment, the material changes colour and becomes magnetic~\cite{23c1640}. 
Olivines annealed in other atmospheres such as \ce{N2} and \ce{Ar} or at lower temperatures in \ce{O2} do not become magnetic (or a change in magnetization is much weaker). The motivation of this study was to better understand the high-temperature annealing (HTA) effects of material changes leading to a change of magnetization in natural olivines harvested from a quarry from an old volcano site in Mortlake, Victoria, Australia. Olivines from this site are known for a darker green colour, which signifies a Fe-rich phase (a lower melting temperature at $\sim 1200~^\circ$C). Magnetisation changes and determination of the Curie temperature of Fe-oxides during HTA is one of the most widely used methods in multi-mineralogical samples~\cite{multi}. 

In previous polarised FTIR measurements~\cite{23c1640}, it becomes clear that samples have to be polished for measurements in reflection. Micro-scale heterogeneity is also challenging to determine composition and orientation using Raman scattering~\cite{Kuebler2006} when large volumes (mm$^3$) have to be characterised. Moreover, the natural olivines have various inclusions such as Cr-rich spinels, garnet, and pyroxenes rich in \ce{SiO2} polymorphs. How to analyse such complex samples using averaging of large data sets and still be representative of the used HTA protocols is the motivation of this study. Figure~\ref{f-oliv} shows the challenge one faces when measuring FTIR spectra of complex heterogeneous samples (Fig.~\ref{f-oliv}(b)), which show location-dependent orientational changes of particular absorbing species as revealed by the four-polarisation (4-pol.) analysis (Fig.~\ref{f-oliv}(c)). 
      
Here, we reveal structural changes of olivine after annealing at temperatures up to 1500$^\circ$C using polarised FTIR reflection spectroscopy at the IR chemical fingerprinting spectral window (2D characterisation of polished samples), X-ray energy dispersive spectroscopy, and magnetization analysis. The program to work with big data sets for multi-parameter analysis: polarization, image site correlation, and samples processed at different HTA temperatures was developed based on 4-pol. method~\cite{4pol,25cbm110573}, which was shown previously to have a super-resolution capability~\cite{19n732,25nse70099}. Any other parameter can be added to trace structure (optical fingerprint) changes on a particular treatment. The IR chemical fingerprinting window is used for structure recognition of various polymorphs in natural heterogeneous samples of olivines (Mortlake, Victoria, Australia) before and after high-temperature annealing. A method is introduced to define synthetic colour (wavelength and bandwidth) with the assignment of RGB bands following an intuitive visible spectral conventions. Such RGB colour changes are due to the HTA and map the corresponding structure changes.     

\section{Experimental: samples and methods}

\subsection{Olivine samples}

The olivine geodes samples were collected at a quarry (extinct volcano) at Mortlake, Victoria, Australia. The geodes were crack-opened, and green olivine interior clumps of crystallites (2-4 cm in cross-section) were collected (Fig.~\ref{f-anne}(a)). These olivines were annealed at temperatures of: 300, 500, 700, 900, 1100, 1300, and 1500$^\circ$C in an \ce{O2} flow of 20~mL/min and 200~mL/min (NABETHERM tube furnace). Also, reference samples were annealed in \ce{N2} and \ce{Ar} flow. 
The HTA caused colour change of samples, and they became magnetised once $\sim 1200^\circ$C temperature was exceeded~\cite{23c1640} (Fig.~\ref{f-anne}(b)). Changes of the HTA-treated olivines had no sign of glassy formations and reflows, only colour and transparency changes linked to phase transformations (Fig.~\ref{f-anne}(c)).

The olivine samples were placed into acrylic resin moulds with Struers epoxy resin (90~g of the EpoFix Resin mixed gently with 10.8~g of the EpoFix hardener) under vacuum for 15~minutes to remove excess bubbles, the sample was set and removed from the mould after 24~hours. It was sliced to expose the cross section of each olivine sample with a 6-inch diamond Wafering Blade (WB-0065HC, PACE Technologies) on a Precision Sectioning Machine (IsoMet 1000, Buehler), the speed was set at $\sim$150 $\pm$ 50~rotation per minute (rpm). Then an automatic grinding and polishing machine (Tegramin-25, Struers) was used for further processes. The polishing protocol consisted of grinding with finer silicon carbide papers ($\#$1200 and $\#$2000, Struers; Plate: MD-Gekko, Struers; speed: 150~rpm; applied pressure: 15 N) 2.5 min per each side, followed by polishing with water-based diamond suspensions (DiaPro Allegro/Largo9 $\mu$m and DiaPro Dac 3 $\mu$m, Struers) for 5.5~min and 4~min respectively per side, then finish the process with oxide polishing suspensions (OP-S, Struers; Plate: MD-Chem, Struers; speed 150~rpm; applied pressure 10~N) for 1.5~min which allows the surface to be $\sim 0.2~\mu$m roughness $R_a$ (shown in Fig.~\ref{f-oliv}(a)). Figure~\ref{f-poli}(a) shows the samples after the annealing, cutting, and polishing in an epoxy puck. Flat polished surface facilitated detailed optical inspection of samples in optical transparency regions and revealed complex patterns of phase transformations, grains, micro-veins, and inclusions such as empty as well as quartz in-grown cavities (Fig.~\ref{f-poli}(b,c)). Distinct features of dendritic formations, recognizable as dark (red-brown to black) structures, were typical of the samples annealed at the highest temperature $\sim 1500^\circ$C (Figure~\ref{f-poli}(c)). 

\subsection{FTIR spectroscopy}

The cut and polished olivine samples were analysed using the FTIR HYPERION 3000 Microscope in reflection mode with a 30~$\mu$m pinhole 
at the infrared microscopy (IRM) beamline of Australian Synchrotron (Fig.~\ref{f-oliv}(a)). The analysis was carried out using 4 polarisations 0$^\circ$, 45$^\circ$, 90$^\circ$, and 135$^\circ$ of incident light and with no polarisation analysis after the reflection. 
The numerical aperture of the 20$^\times$ magnification Cassegrain objective was $NA$ = 0.60, which defined the diffraction limit of $\lambda/NA= 16.7~\mu$m for $\lambda = 10~\mu$m (wavenumber $\Tilde{\nu} = 1000~$cm$^{-1}$). Note: we use $\lambda$ and $\Tilde{\nu}$ interchangeably to represent wavelength and wavenumber, respectively, throughout analysis and discussions below. 

Data was collected using the OPUS program. For multi-parameter spectral analysis of mapped regions at different polarisations and samples processed at different HTA temperatures, a software platform was created in this study.

\subsection{Spectral analysis tool}

For multi-parameter (polarisation, temperature) data analysis at different locations of the same sample and a series of samples annealed at different temperatures, a program was created/developed. A new feature for data analysis was introduced for selective assignment of up to three spectral windows customarily named as RGB following visible colour conventions (Fig.~\ref{f-soft}; real spectra of a 1500$^\circ$C annealed sample of olivine are used). Such an assignment allows to select specific absorbance or reflectance bands ($A$ or $R$) for the identification of a specific phase of material (e.g., an olivine phase and to distinguish it from other phases present in the sample). All mapped data points (a matrix of $10\times 15$ pixels) can be analysed for the presence of specific RGB crystalline absorption bands (Fig.~\ref{f-soft}(a)). For a selected pixel, interrogation of the local orientation of the crystallite (its absorption band) can be made using 4-pol. method~\cite{4pol}. For a linearly polarised incident light defined by its azimuth orientation angle $\theta$, four measurements are carried out at $0,45,90,135^\circ$. The azimuthal dependence of an absorber $\theta_{A}$ can be determined~\cite{4pol}:
\begin{equation}\label{e-A}
 A(\theta) = \frac{A_{max} - A_{min}}{2}\cos(2\theta - 2\theta_{abs}) + \frac{A_{max} + A_{min}}{2},
\end{equation}   
\noindent here the absorbance $A$ is determined from the transmittance $T$ as $A = -\log_{10}T$; it can also be useful to use a fit by a single-angle function $\cos^2\theta = (1+\cos{2\theta})/2$ as in Fig.~\ref{f-oliv}(c) to reveal the orientation of absorbers.

In this study, the 4-pol. method for the determination of the IR absorption bands out of the reflectance spectra is introduced. The distinctive reflectance $R$ features can be selected to reveal their orientation as shown in Fig.~\ref{f-oliv}(c). For the absorbance determined from transmittance, it is possible to use the analytical Mueller matrix approach~\cite{23c1640} to determine polarisation changes of transmitted light, however, it is not directly applicable to measurements in the reflection imaging mode. 

Figure~\ref{f-poli}(b) shows a user interface snapshot for the local single pixel analysis using 4-pol. method. At a selected wavelength, e.g., corresponding to the known absorbance band of olivine, the fit by Eqn.~\ref{e-A} determines the orientation of absorbers $\theta_A$ (maximum absorbance) and $A_{max,min}$ values. The fit by Eqn.~\ref{e-A}: $A$ vs. $\theta$ can be conveniently change to $A$ vs. $\sin\theta$ (inset in (b)). For these three parameters to be determined, at least three polarisation angles of $A$ spectra are required as previously shown~\cite{23c1640}. Use of 4 orientation angles makes it convenient to express $Amp = A_{max}-A_{min} = \sqrt{(A_0-A_{90})^2 + (A_{45}-A_{135})^2}$ from four measured absorbance values $A_{0,90,45,135}$ and to determine orientation azimuth of absorbers $\theta_A = \left[\tan^{-1}{\frac{A_0-A_{90}}{A_{45}-A_{135}}}\right]/2$~\cite{4pol}. Obviously, more measured $A$ (or $R$) points at a larger number of $\theta$ orientations of the linearly polarised incident light can be used for a better fit (not utilised in this study due to long collection time per spectrum in the hyper-spectral FTIR imaging).


\subsection{Characterisation of magnetic properties}

Magnetisation of pristine olivine (RT) and HTA treated at different conditions was determined using the Bartington MS2 Magnetic Susceptibility Metre equipped with MS2K-probe. The probe provides surface measurements across a $\sim 25$~mm diameter area with sensitivity that retrieves 100\% of the magnetic signature from the first millimetre and up to 50\% of the signature at 3~mm of penetration. The smallest measurement achievable by the probe was $1\times 10^{-6}$ in SI units~\cite{Andy2013}.


\subsection{Structure-composition characterisation}

Scanning electron microscopy (SEM) was carried out with a JEOL JSM-IT800 equipped with X-ray energy dispersive spectroscopy (EDS) analysis. An accelerating voltage of 10~kV was used, at a working distance of 9.6~mm. Point EDS analysis and element colour mapping were carried out with the AZtec software package (Oxford Instruments). 

\begin{figure*}[t]
\centering\includegraphics[width=16cm]{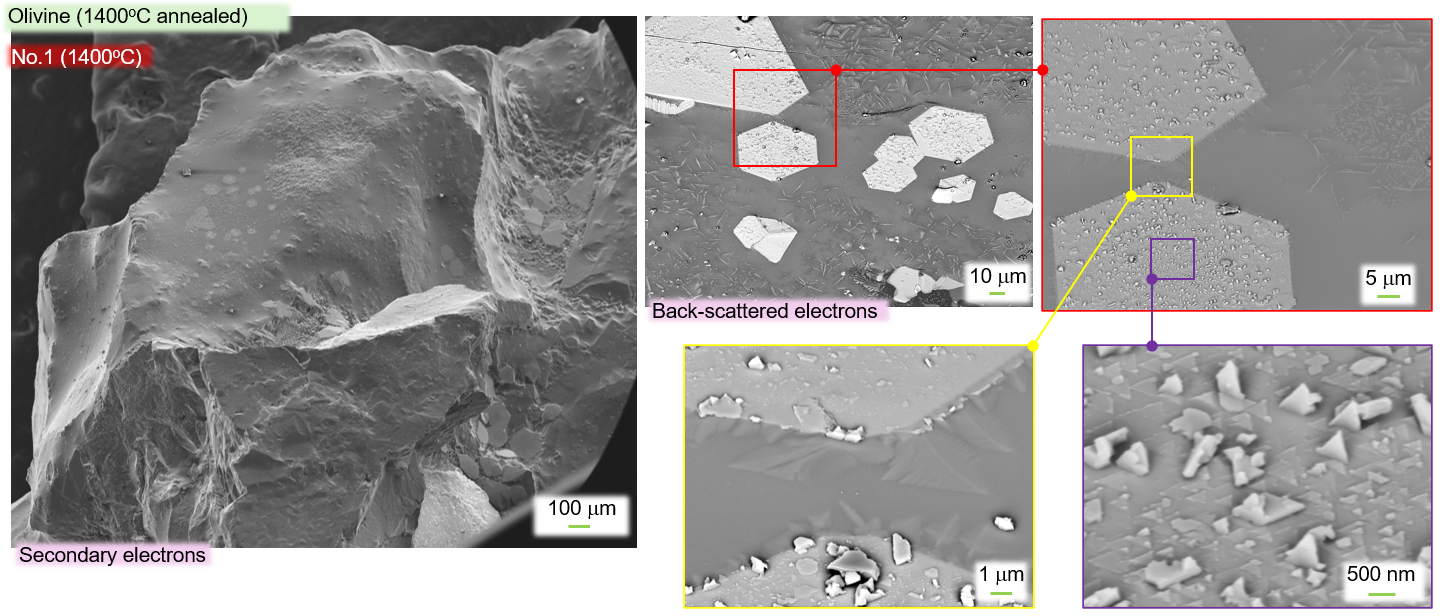}
  \caption{SEM image (back-scattered and secondary electrons) of an olivine (Mortlake) grain annealed at 1400$^\circ$C [in 
 air for 2~hours]. Close-up views of the crystalline formations used in EDS analysis. Dendritic formations are recognisable as those in optical images (Fig.~\ref{f-poli}).     }\label{f-1400}
\end{figure*}
\begin{figure*}[tbh]
\centering\includegraphics[width=16cm]{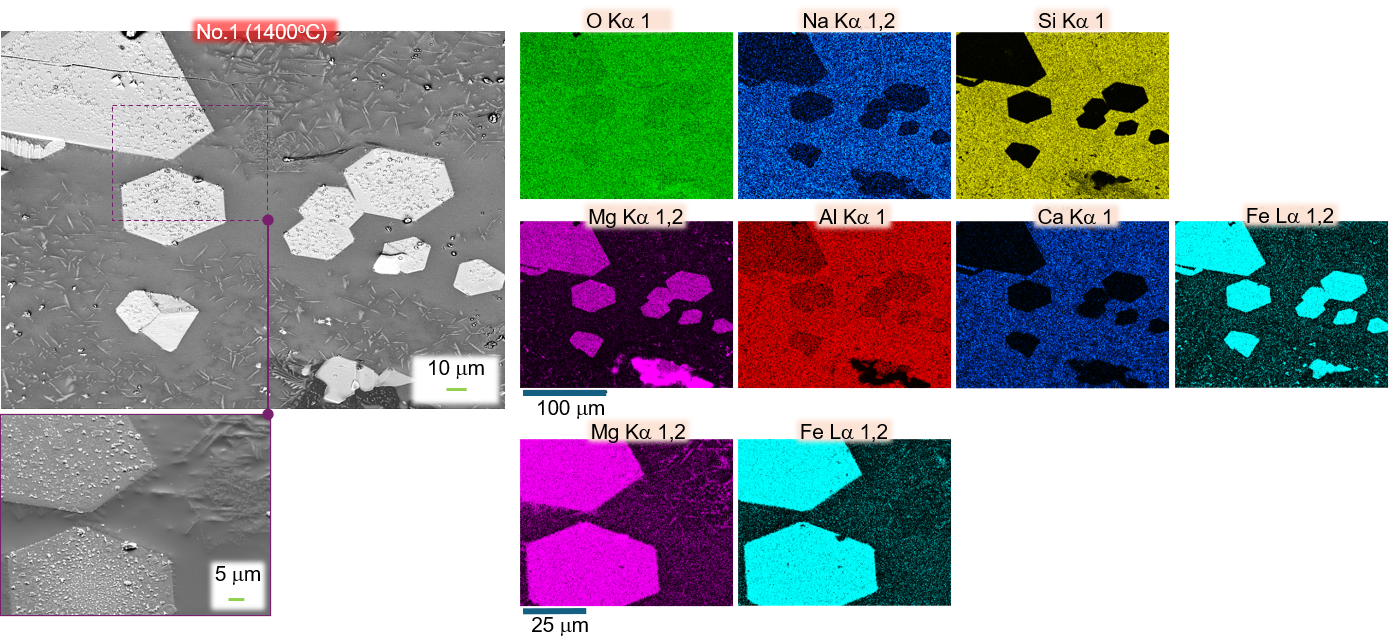}
  \caption{EDS an olivine grain annealed at 1400$~^\circ$C (same as in Fig.~\ref{f-1400}). Elemental maps for O, Na, Si, Mg, Al, Ca, Fe for K$_\alpha$ and L$_\alpha$ shell transitions from SEM images at different magnifications. Formation of Mg- and Fe-enriched crystals is evidenced (only Mg and Fe selected for the bottom row). }\label{f-EDS}
\end{figure*}
\begin{figure*}[tbh]
\centering\includegraphics[width=16cm]{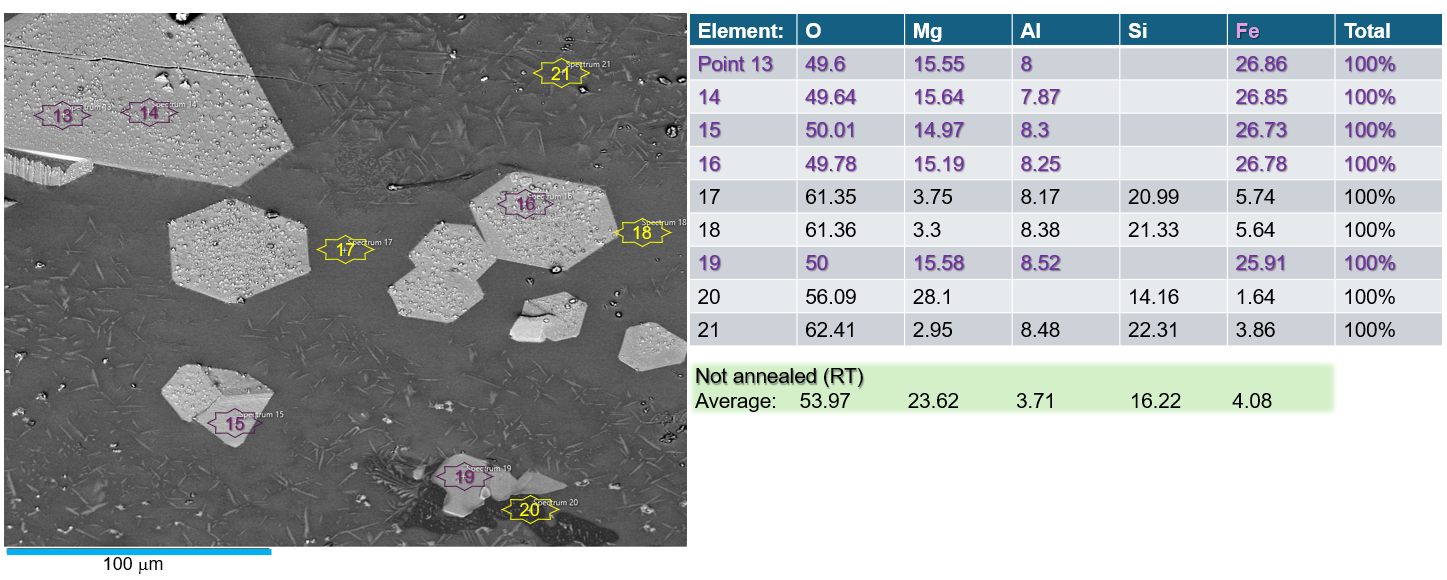}
  \caption{Quantified EDS analysis from the selected regions on olivine annealed at 1400$~^\circ$C; the area of single spot analysis was $\sim 0.25\times 0.25~\mu$m$^2$. Same sample as in Fig.~\ref{f-1400}; the marker is centered at the point of analysis. Fe enrichment with simultaneous Si depletion is observed on the crystal sites. See Supplement for the reference not annealed (RT) olivine Sec.~\ref{RT}. }\label{f-Fe}
\end{figure*}
\begin{figure*}[tbh]
\centering\includegraphics[width=17cm]{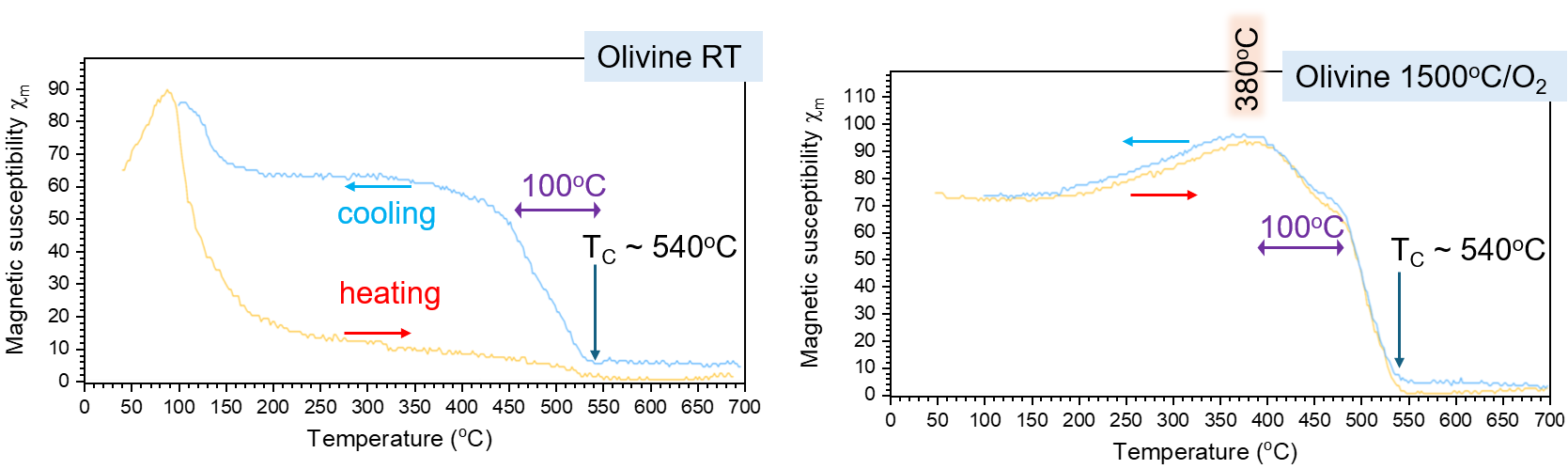}
  \caption{Magnetic volume susceptibility $\chi_m$ 
 of olivines: pristine (RT) and HTA at 1500$^\circ$C in \ce{O2}; for diamagnetic $\chi_m < 0$, e.g., forsterite \ce{Mg2SiO4} is diamagnetic. Heating was up to $700^\circ$C during magnetisation measurements; heating rate was 10$^\circ$C/min. An approximate amount of micro-granular olivine was $\sim$5~g. }       \label{f-magn}
\end{figure*}

\section{Results}

To determine the effect of HTA on structural change of olivine samples (see plethora of possible phase transitions in Fig.~\ref{f-min}), the spectral region with major reflectance $R$ features from 820~cm$^{-1}$ to 1020~cm$^{-1}$ is useful. Figure~\ref{f-oliv}(b) shows 
4-pol. discriminated reflectance $R$ spectra (with 45$^\circ$ linear polarisation shift) from all RoI (a) in a pristine (room temperature RT) sample of polished olivine. Optical image at visible range shows a characteristic vein pattern on olivine. The $R$ spectra presented in (b) are the total average from the measured points in RoI. Even for the averaged data over $10\times 15$ measured points, characteristic $R$ peaks show distinct polarisation dependence (c); note, absorbance maxima would correspond to the valleys in the spectra. The validity of such an averaging approach, illustrated in Fig.~\ref{f-oliv}(b), where all measured points for ROI are summed, can be used to establish global trends of structural changes due to HTA treatments of olivines, changes of crystalline phases, etc. The peak with two side lobes within 980~cm$^{-1}$ to 850~cm$^{-1}$ is characteristic to fayalite and has strong orientation dependence~\cite{fayalite} (Fig.~\ref{f-oliv}(b)). Upon HTA, it changed to forsterite \ce{Mg2SiO4} (for both, experiments were carried out with and without \ce{O2} flow). From 300$^\circ$C upwards, the uniform peak splits into two peaks and begins to merge at the higher and higher temperatures. At 1500$~^\circ$C, the peaks have a very complex pattern as could be inferred from Fig.~\ref{f-soft}(a). Shift of IR absorption peaks toward longer wavelengths (lower frequencies) with simultaneous peak broadening is due to formation of Fe-rich olivine and a greater ionic mass (Fe content increases from forsterite to fayalite).

Determination of local and especially global (averaged over that mapped region) orientation of olivine and its different phases and newly formed crystalline phases due to HTA requires a big data approach developed for this study, as discussed next.


\subsection{``RGB'' analysis at IR range }

To determine global trends in structural changes during HTA treatment of different samples, a big data approach was implemented, i.e., a database of point-by-point spectra was collected from samples experiencing different HTA temperatures and was analysed at specific spectral bands. For intuitive convenience, we determine them as RGB for selected regions. Figure~\ref{f-soft}(a) shows RGB fictive-colour assignments and all family spectra from the scanned region with sub-1~mm cross section. The relative weight of RGB components defines the colour of the spectral line and region on the map. 
This visualisation is helpful to recognise global trends of IR absorption in different locations as well as their orientation at the selected wavelength (colour) as shown in Fig.~\ref{f-soft}(b). An extensive family of experimental data is shown in the Supplement in Figs.~A8-A18 for olivine samples annealed at different temperatures. After HTA, samples were cut through the middle and polished for IR reflectance measurements. This preparation mode allowed inspection of the internal rearrangement of composition, which can differ from EDS analysis on the surface of the sample after HTA. If reflectance is changed due to absorbance in sub-surface depth of a few microns $\sim\lambda/4$, the experimental spectra of reflectance resemble those of absorbance; the spectra were treated as absorbance for the analysis.      


The colour maps in Figs.~A8-A18 reveals tendency of absorption peak widening with high temperature of HTA. The influence of ambient gas \ce{O2} vs. \ce{N2} was negligible, understandably due to a large volume of sample which was inspected after cutting through its middle. The presence of fosterite and fayalite, sharper and broader absorbance peaks, respectively, was in all samples before and after HTA. Since different samples were used from the same larger regolith, the heterogeneity of the IR spectra is large. The HTA was taking a couple of hours, which is a short time for ion diffusion, which becomes more prominent only at $>1100^\circ$C in olivine with a coefficient $10^{-16} - 10^{-18}$~m$^2$/s and an Arrhenius type of activation energy $\sim 2$~eV/atom ~\cite{diff,dif1}. The fastest diffusion is along [001] c-axis by a factor of 10-100 times as compared with other directions~\cite{dif1}. Strong anisotropy in new phase precipitation is expected upon HTA of olivine. 

By inspection of Figs.~A8-A18, it is evident that the boundaries between different component phases are revealed in colour maps. Defect lines and micro-cracks sometimes exist in the same phase (optical image) and are confirmed by the colour maps.  


\subsection{X-ray Energy Dispersive Spectroscopy (EDS) }

Spectroscopic image analysis was carried out on flat, mirror-polished samples, which were sliced after HTA treatment. This was required for reliable 4-pol. analysis in reflection. For EDS, we used olivine grains without polishing to avoid contamination. Since atomic composition analysis is carried out with an electron beam, there was no necessity to have a polished sample.  

Upon HTA of olivine, a strong colour and structural changes were observed inside sliced samples by optical micro-imaging (Fig.~\ref{f-poli}). Specific dendritic darker regions were more pronounced after HTA at higher temperatures. The composition change of olivine upon HTA was inspected by X-ray EDS from SEM imaged micro-regions on a 1400$^\circ$C annealed olivine sample (Fig.~\ref{f-1400}). Patches of crystalline formations were discernible on the surface with dendritic structures at their rims. Also, 3D nano-structures were formed on the surface as revealed by SEM. Atomic composition maps were obtained from those regions (Fig.~\ref{f-EDS}). The crystalline regions showed increased content of Mg and Fe with clear depletion of Si as well as Na, Al, Ca. This clearly shows decomposition of olivine \ce{(MgFe)2SiO4}, where Si is a major component, which is absent at the HTA crystallised regions. By a point-site EDS analysis, the atomic ratios were obtained at several specific locations (Fig.~\ref{f-Fe}). The largest change from the non-annealed (RT) olivine was a large content of Fe (a change from $\sim 4\%$ to $\sim 26\%$) while Si was depleted from 16-20\% to an undetectable amount (see table in Fig.~\ref{f-Fe}). This qualitatively confirms the formation of Fe-rich oxides/composites, which rendered the material magnetic after HTA. 

The following spinel group minerals of the structure \ce{XY2O4} could be expected after HTA: spinel \ce{MgAl2O4}, hercynite \ce{FeAl2O4}, pleonaste - a natural solid solution between spinel and hercynite - \ce{(MgFe)Al2O4}, magnesioferrite \ce{MgFe2O4} (magnetic), and magnetite \ce{Fe3O4} (or \ce{FeFe2O4} a iron spinel with one Fe in 2$^+$ and two in 3$^+$ state). Also, formation of ferropericlase (Mg,Fe)O, the second most abundant mineral in the Earth's lower mantle is expected from a solid solution of periclase MgO and w\"{u}stite FeO. All this family of phases has no Si, i.e., no mineral-forming silicates: olivine \ce{(MgFe)2SiO4}, pyroxene \ce{(MgFe)SiO3}, garnet \ce{Fe3Al2(SiO4)3}. The observed ratios of Mg:Fe:O are close to spinel 1:2:4 or 14\%:28\%:58\% (Fig.~\ref{f-Fe}). The clear trend observed in annealed samples was depletion of Si in those regions that were enriched by Fe. It corresponded to the formation of micro-crystallites as a result of HTA.     

\subsection{Magnetisation}

Figure~\ref{f-magn} shows magnetic susceptibility $\chi_m$ vs. temperature cycling from RT-to-700$^\circ$C in olivine samples before and after annealing at 1500$~^\circ$C in \ce{O2}. The Curie temperatures of the Fe-oxides: magnetite \ce{Fe3O4}, maghemite $\gamma$-\ce{Fe2O3}, and hematite $\alpha$-\ce{Fe2O3} are 580$^\circ$C, 300-400$^\circ$C (magnetisation of maghemite is $\sim 150^\times$ greater than that of hematite), and $680~^\circ$C, respectively~\cite{multi}; $T_c = 770^\circ$C for Fe. Maghemite is often a low-temperature oxidation product of magnetite~\cite{multi}, which both could be contributing to the magnetisation of HTA 1500$^\circ$C annealed olivine (Fig.~\ref{f-magn}).
The reversible slope change of magnetisation at $\sim 460^\circ$C is close to the $T_C = 465^\circ$C of \ce{MgFe2O4} (formed upon heating from \ce{MgO} and \ce{Fe2O3})~\cite{complex}. 

The FeO can also be contributing to phase changes, especially in \ce{O2} environment as widely researched and relevant to the batteries~\cite{Zhao2013}. The Fe(II) FeO is thermodynamically unstable below 575$~^\circ$C and disproportionates to Fe and \ce{Fe3O4}, both ferromagnetic.

Magnetic properties of other related iron compounds are reviewed in ref.~\cite{multi}, with the most relevant solid solutions with Ti and Cr. In those cases, similar tendencies as for maghemite-to-(magnetite, hematite) transitions take place in the window of temperatures 350-550$~^\circ$C with $T_c$ of magnetisation disappearance at 550-600$~^\circ$C. 


\section{Discussion}\label{disco}

Analysis of local orientation and crystalline phase determination using chemical maps introduced in the newly developed online application for IR chemical fingerprinting spectral range can be extended to shorter visible as well as to longer THz spectral ranges. It is important that the reflective Cassegrain optics used in this study provide equivalence in transmittance $T$ and reflectance $R$ measurements without alteration of linear polarisation, e.g. due to birefringence. This property is not straightforward when a half-mirror at $45^\circ$ incidence (e.g., 10 nm gold-coated coverglass) is used in $R$ imaging at the visible spectral range. The reflection and transmission of linearly polarised light in the case of 10-nm-thick Au films have to be properly accounted for the phase change before and after the sample. In the case of fully reflective objective lenses (no dichroic nor half mirrors), there was only an orientation shift introduced by a different number of steering mirrors towards the sample (Fig.~\ref{f-pol}), however, this is accounted by normalization to the reflectance of the gold mirror.

Reflectivity at IR at the tailored RGB bands representative to specific phase (in the sample) can be useful in multi-phase and heterogeneous samples, as in this study of olivines. Other optical techniques at the visible spectral range, such as absorption spectroscopy, Raman scattering, and photoluminescence, can, in principle, characterise material from even smaller focal volumes. Olivines with a large bandgap (8.4~eV for forsterite~\cite{bandgap}), large birefringence (biaxial crystal), a large number of impurities, and complex composition (natural crystals)~\cite{complex}, defects on a micrometer scale~\cite {defects} make those measurements more demanding for larger samples, larger processed areas/volumes (mm-to-cm cross sections).   

\section{Conclusion and outlook}

The IR chemical fingerprinting spectral region for structural analysis was used with synthetic RGB band assignment to visualise/trace phase changes introduced by HTA. Spectral IR window $\lambda = 8 - 24~\mu$m was utilised for analysis under focusing with $NA = 0.4 - 0.6$ objective lenses to characterise focal spots of 20-100~$\mu$m in diameter (and 50 - 300~$\mu$m depths) point by point. Such sampling is appropriate to characterise global material composition over the selected sampled 0.5-2.5~mm RoIs on the polished natural mineral/rock samples. The correlation of optical micro-characterisation and global changes of natural olivine regolith/xenolith was carried out by atomic analysis (EDS) and magnetisation. Changes introduced via HTA can be spectrally detected in global tendencies of IR spectral band changes and correspond to structure/phase changes of multi-component samples. This synthetic-RGB approach can be extended to add other processing parameters. The virtue of a real-time data analysis toolbox available as freeware online simplifies data analysis and is aligned with further automation of specific tasks. Measured IR spectra are uploaded as a database for cross-correlations. A useful feature is that the optical thumbnail image of RoI is also available to investigate specific regions, e.g., defects, grain boundaries, etc. Correlation of optical colour images with IR spectral properties together with polarisation information could contribute to large-scale analysis of material~\cite{red}.    

In future studies, the Principal Component Analysis (PCA) can be applied to the experimental spectra for establishing correlations between specific phases and HTA conditions. The presented ``optical biopsy'' can be complementary to the Light Induced Breakdown Spectroscopy (LIBS) in elemental composition from ablation spectra (``real biopsy'') of probed minerals.

\small\begin{acknowledgments}
The synchrotron polarized-FTIR 
experiments were undertaken at the Infrared Microspectroscopy (IRM) beamline at the Australian Synchrotron, part of ANSTO, via a merit-based beamtime proposal (ID. M22505) in 2024. The Far-IR spectral range was achieved using a Si:B photodetector from the THz beamline 
on established (2019) the 4-angle polarization method. 
SJ is grateful for support via the Australian Research Council Discovery DP240103231 grant. JM acknowledges support via JST CREST (Grant No. JPMJCR19I3) and KAKENHI (No.22H02137). MR was supported via KAKENHI (Grant No. 22K14200) and SAKIGAKE (Grant No. JPMJPR250E). 

We are grateful to a volunteer fossicking guide, Joe Mooney from Mortlake, Victoria, for leading our collection of olivine xenoliths from the Mortlake quarry in 2016.
\end{acknowledgments}

\bibliography{oliv}

\appendix
\setcounter{figure}{0}\setcounter{equation}{0}
\setcounter{section}{0}\setcounter{equation}{0}
\renewcommand{\thefigure}{A\arabic{figure}}
\renewcommand{\theequation}{A\arabic{equation}}
\renewcommand{\thesection}{A\arabic{section}}

\section{Throughput of linear polarised synchrotron radiation}

\begin{figure*}[tb!]
\centering\includegraphics[width=15cm]{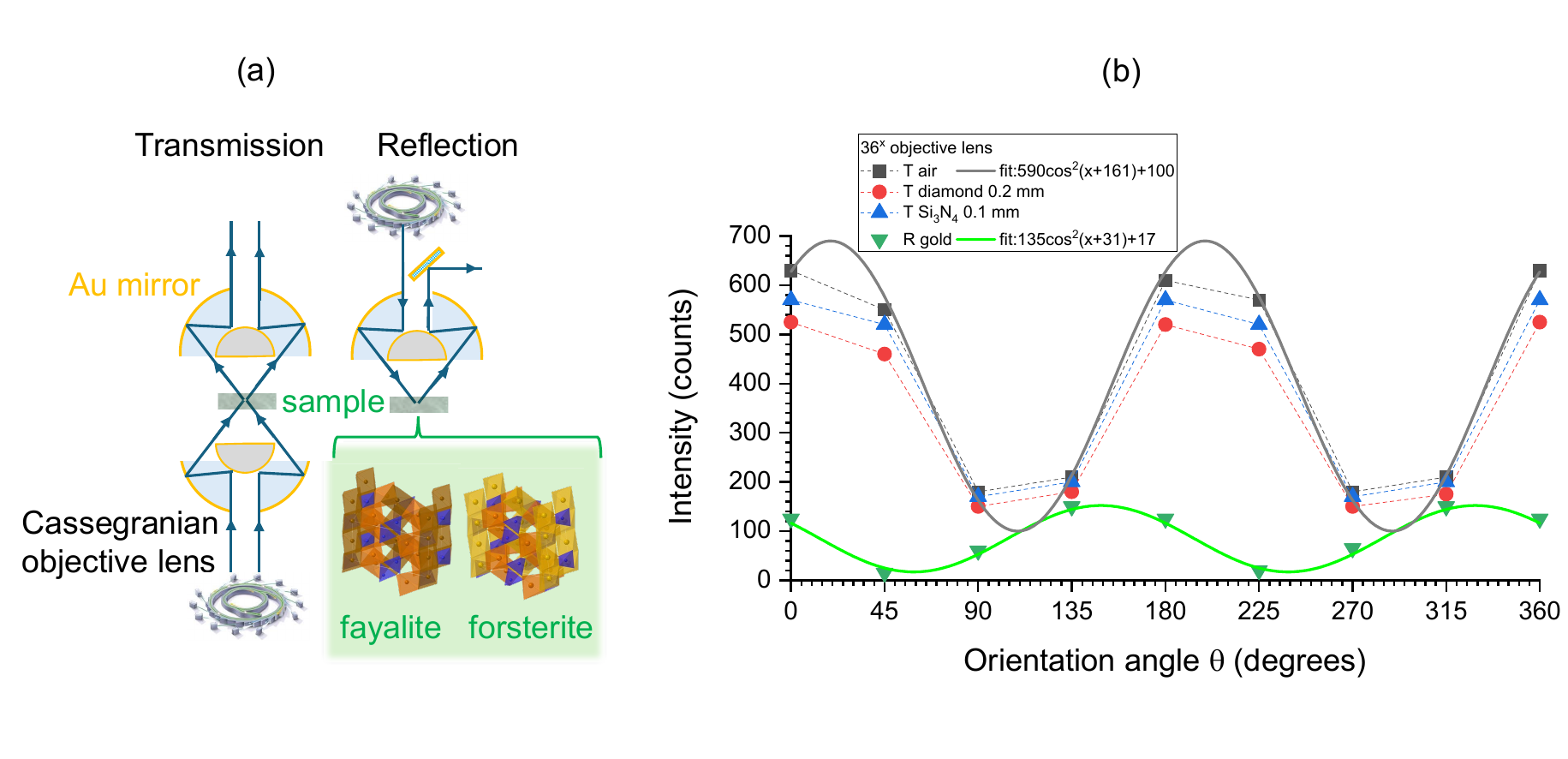}
  \caption{(a) $T$ and $R$ modes of throughput IR signal from synchrotron at the used setup for 36$^\times$ magnification Cassegrainian objective lens of numerical aperture $NA = 0.6$. Samples in this study were olivine family materials (polyhedral structure of two types of olivines is shown (Crystal Maker)). (b) Detected signal $T,R$ vs the orientation azimuth $\theta$ of the wire grid polariser 
 for different samples (air is for the no-sample condition). Dots are experimental data shown with fitted Malus law dependence $\cos^2\theta$. There is a phase shift of 130$^\circ$ (or $-50^\circ$) between $T$ and $R$ modes due to different number of reflections at the mirrors used to guide synchrotron radiation to the sample and detector upon reflection. The phase shift upon normal reflection from Au-mirror is $\pi$ (180$^\circ$) and the difference between s- and p-pol. components is increasingly large at angles of incidence $\geq 45^\circ$.  }\label{f-pol}
\end{figure*}

It was tested that transmitted power of synchrotron radiation was following the Malus law for linearly polarised light at different orientation azimuths (Fig.~\ref{f-pol}). From transmitted power fit without sample (air in (b)) it was possible to determine degree of linear polarisation (along the horizontal fork-mirror used for beam extraction from synchrotron ring) as
$Lin = 590/(590+100)\approx 85.5\%$. The offset of intensity at 100 counts is due to isotopically polarised (radial) component of radiation due to simultaneous contributions of dipole and edge radiations~\cite{18jo035101} while the linear (horizontal $\theta = 0^\circ$ polarisation) is alternating from max-to-min transmission over $\theta$-dependence.  

\begin{figure*}[tb!]
\centering\includegraphics[width=11cm]{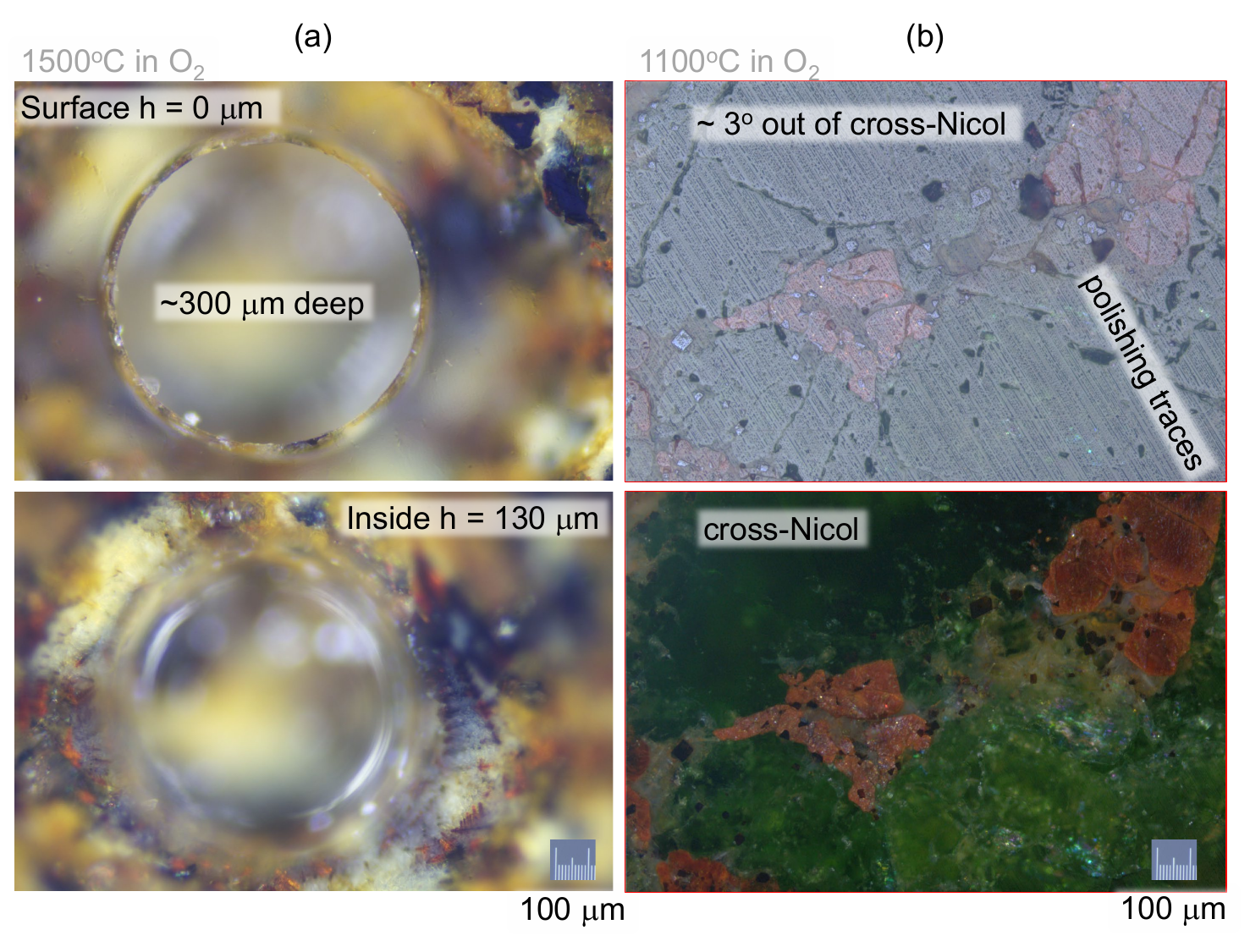}
  \caption{Optical reflection images of olivine polished samples HTA in \ce{O2}. (a) Hemispherical cavity at the surface and at 130~$\mu$m depth revealing dendritic structures; $NA = 0.5$, HTA at $1500^\circ$C. (b) Cross-polarised image at slight mis-alignment and at fully cross position at the same long acquisition time; HTA at $1100^\circ$C. Pattern of dark domain boundaries is clearly discernable and corresponded to the bright-field image on FTIR microscope, see RoI in Fig.~\ref{f-oliv}(a).    }\label{f-dend}
\end{figure*}

\section{EDS from room temperature olivine}\label{RT}

\begin{figure*}[tb!]
\centering\includegraphics[width=16cm]{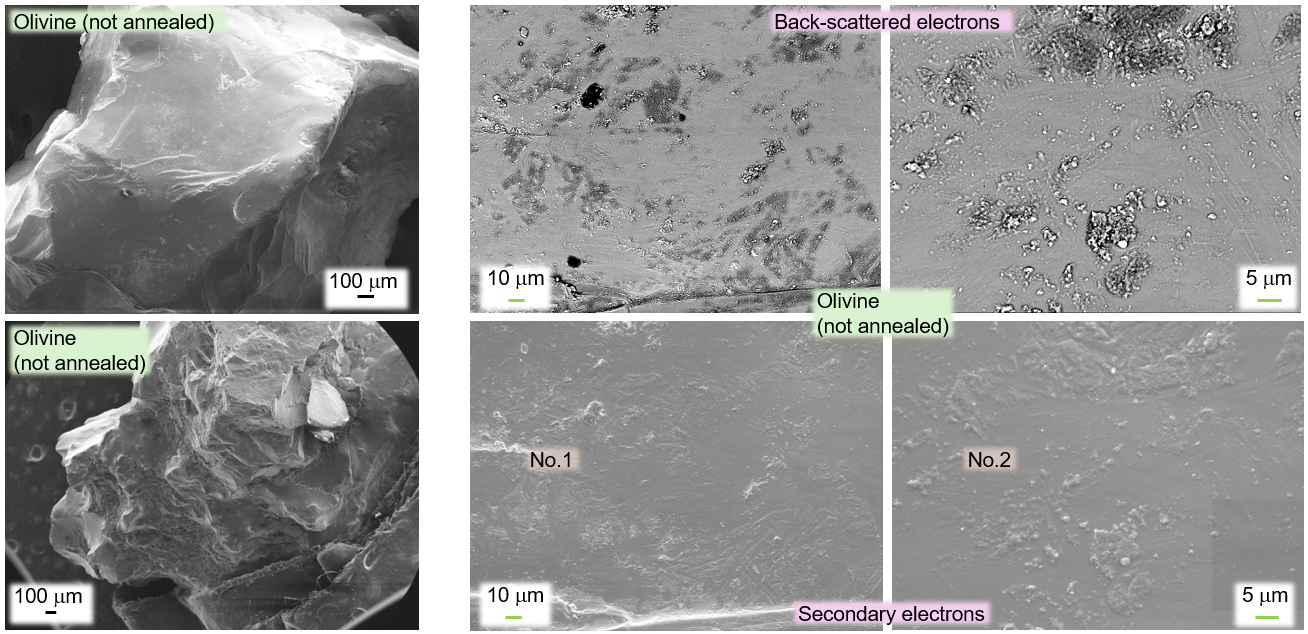 }
  \caption{SEM images (back-scattered and secondary electrons) of a pristine olivine (Mortlake) grain. Close up views at the locations used in EDS analysis.  }\label{f-RTse}
\end{figure*}

\begin{figure*}[tb!]
\centering\includegraphics[width=17cm]{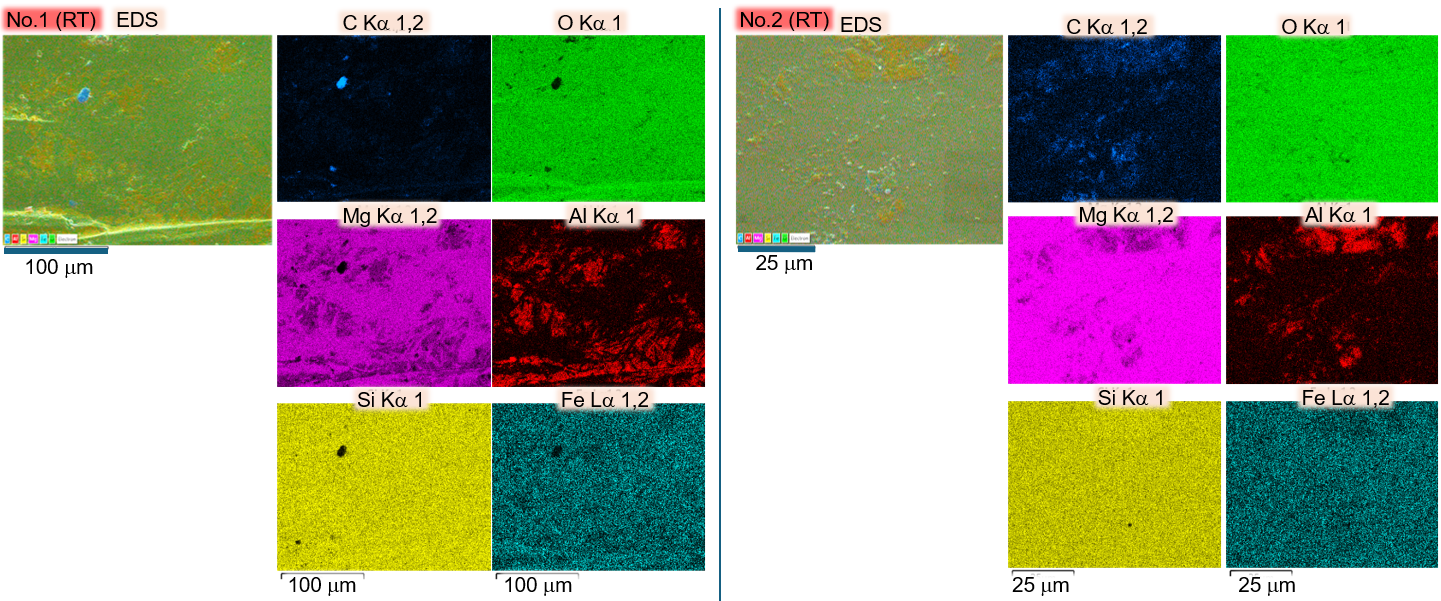  }
  \caption{EDS from two regions of an olivine grain (not annealed). Elemental maps for C, O, Si, Mg, Al, Fe for K$_\alpha$ and L$_\alpha$ shell transitions from SEM images shown in Fig.~\ref{f-RTse}.   }\label{f-RTma}
\end{figure*}

\begin{figure*}[tb!]
\centering\includegraphics[width=14.5cm]{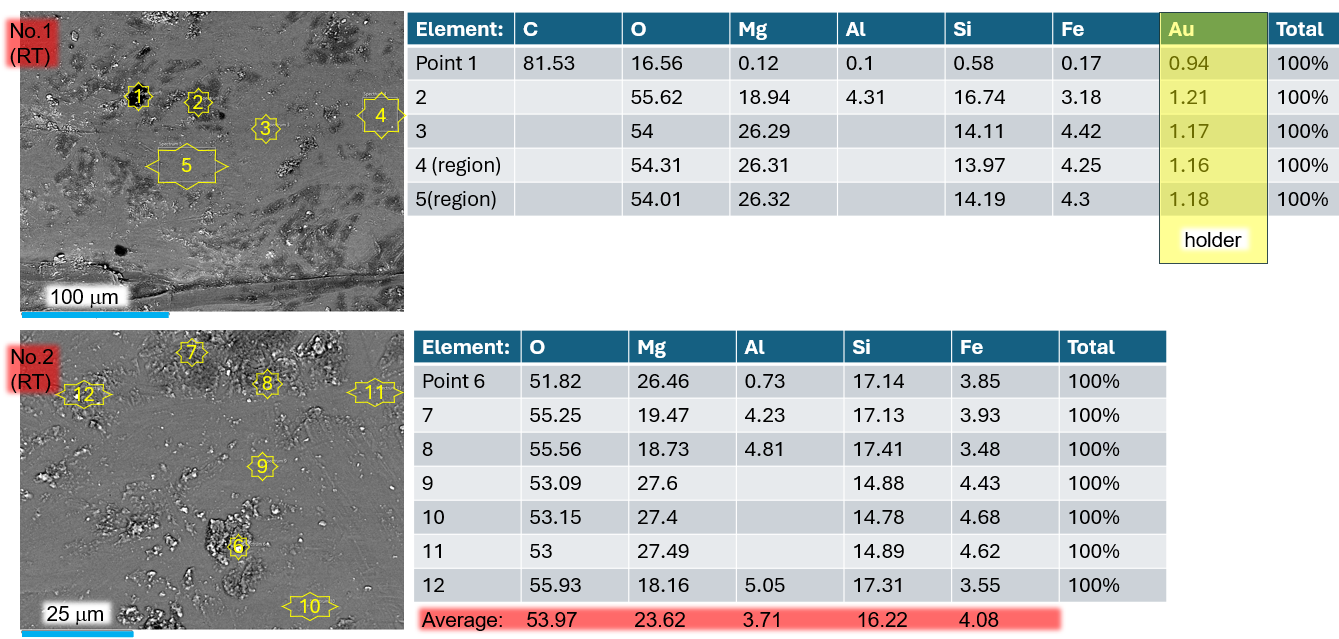  }
  \caption{Quantified EDS analysis from the two selected regions on pristine sample of olivine; the marker is centred at the point of analysis or is enclosing the entire scanned region (No. 4 \& 5). Larger markers indicate correspondingly larger area of analysis while the smallest markers were from $\sim 5\times5~\mu$m$^2$ areas. Atomic Fe concentration was low $\sim 4$. Gold background was from the sample holder.   }\label{f-RTra}
\end{figure*}

The EDS analysis of reference sample - pristine olivine (not annealed) - was carries on a typical green olivine grain. Its SEM images are shown in Fig.~\ref{f-RTse}. Sample was rinsed in IPA before SEM and EDS. Heterogenous atomic distributions of elements recognisable on the surface is shown in the corresponding maps for two samples in Fig.~\ref{f-RTma}. Quantified EDS atomic ratios are shown in Fig.~\ref{f-RTra} with numerical values summarised in the tables.

\begin{figure*}[tb!]
\centering\includegraphics[width=18cm]{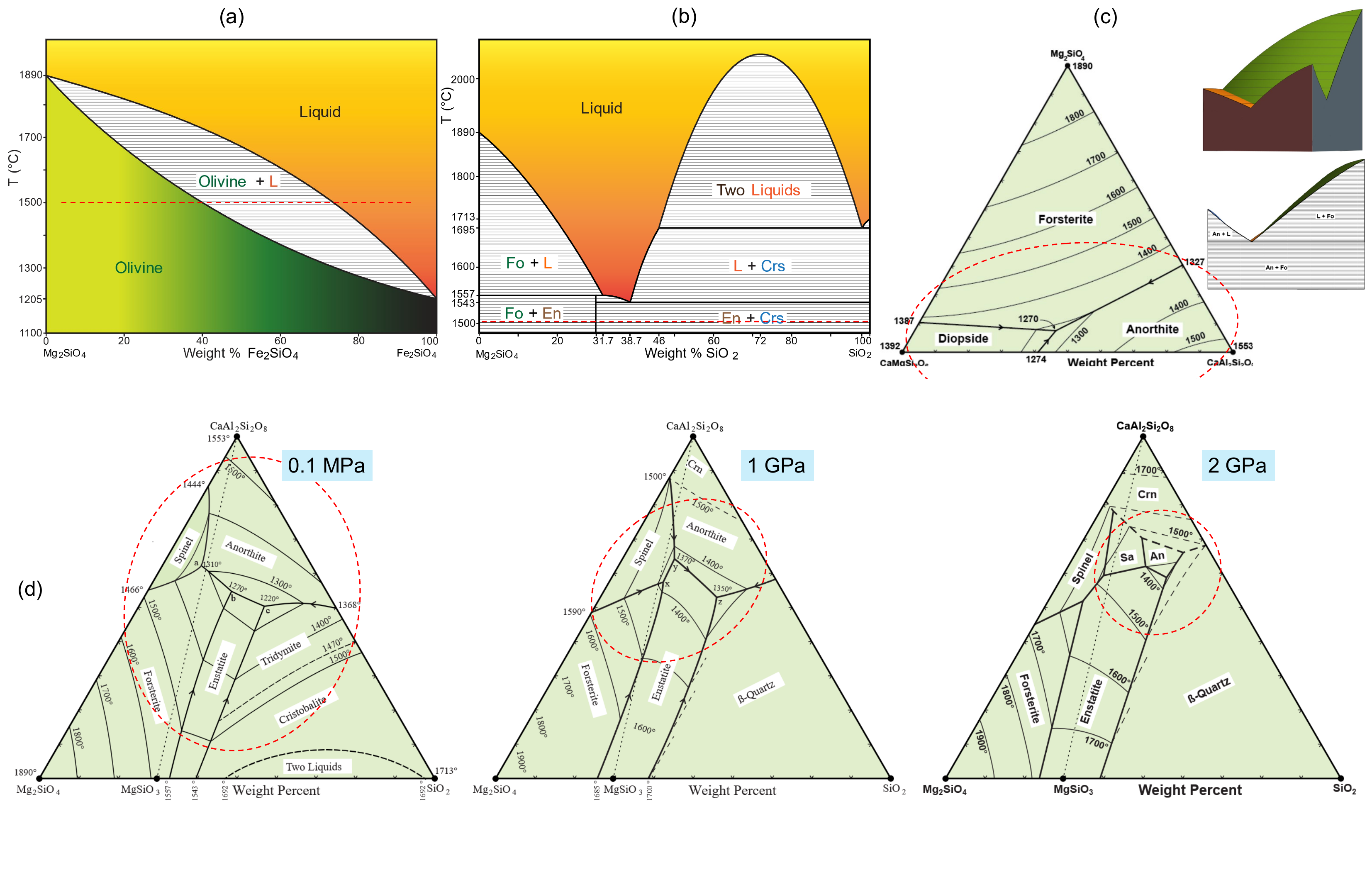}
  \caption{Phase diagrams of olivines and related materials [https://www.science.smith.edu/~jbrady/petrology/front-matter/contents.php; visited 7.3.2026]. Red-dashed lines marks maximum temperature used in this study. (a) Olivines: Forsterite-Fayalite, (b) Forsterite-Silica. Trinary phase diagrams: (c) Diopside-Forsterite-Anorthite, (d) Forsterite-Anorthite-Silica with Spinel at different pressures 0.1~MPa, 1~MPa and 2~GPa.    }\label{f-min}
\end{figure*}
\begin{figure*}[tb!]
\centering\includegraphics[width=14cm]{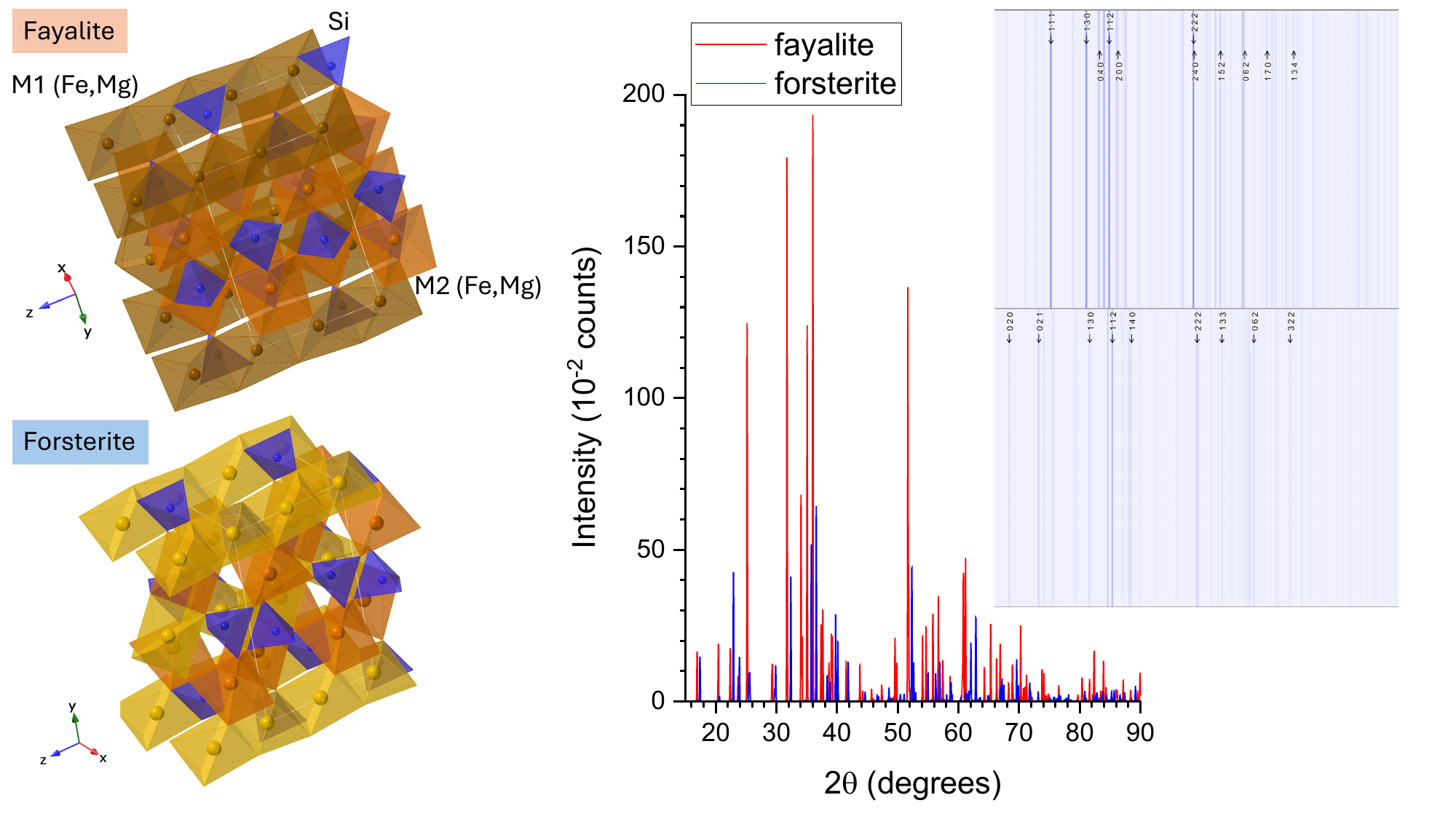}
  \caption{Structures and XRD of Fayalite and Forsterite (CrystalMaker, CrystalDiffract).   }\label{f-xrd}
\end{figure*}


\section{Maps of olivine samples}\label{Samples}

Images of olivine samples annealed at different temperatures (RT - 1500$^\circ$C) in air, \ce{O2}, and \ce{N2} flows are shown collected (Figs.~A8-A18). The intensity scale for all the maps is the same for relative comparison. Colour coding of spectral lines and location on the images are obtained from the relative intensities of the contribution in the assigned RGB bands (according to their spectral location and width). The optical image is overlaid with the scanned region. The pixels' colour is corresponding to the spectral line. When measured location is on the lines between different sample's regions, only intensity is lower, but the colour is the same. 

The pixelated colour maps for each of four polarisations makes clear visualisation if colours are changing (not intensity). Usually those regions are separated by the apparent veins or defect lines which demarcate different regions in the sample. Colour change with polarisation signifies presence of dichroism (see, e.g., Fig.~A14).  
\begin{figure*}[htbp]
  \centering
  \includegraphics[width=0.89\textwidth]{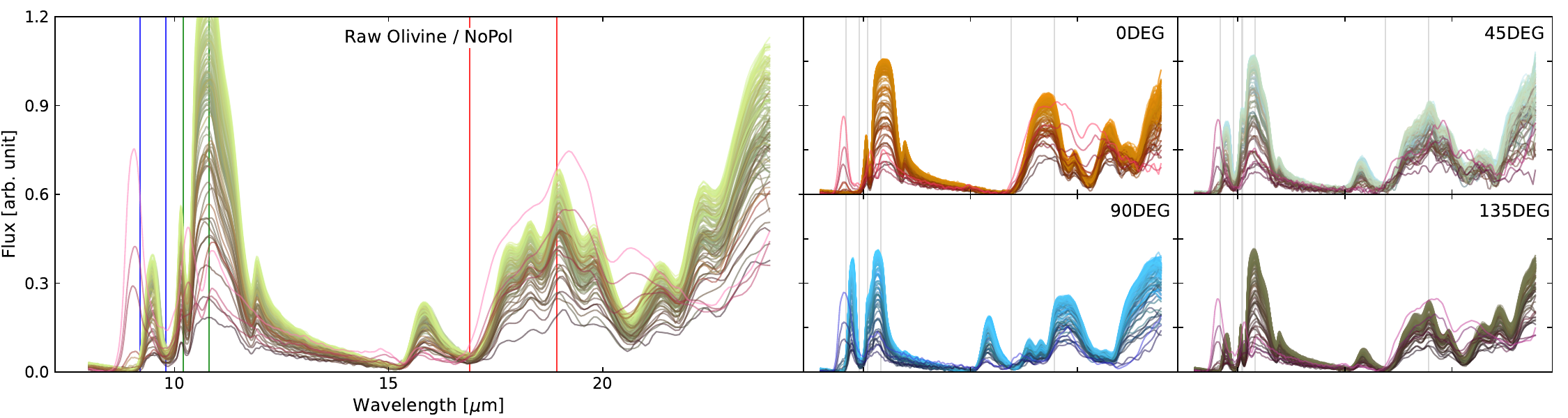}
  \hfill
  \includegraphics[width=0.89\textwidth]{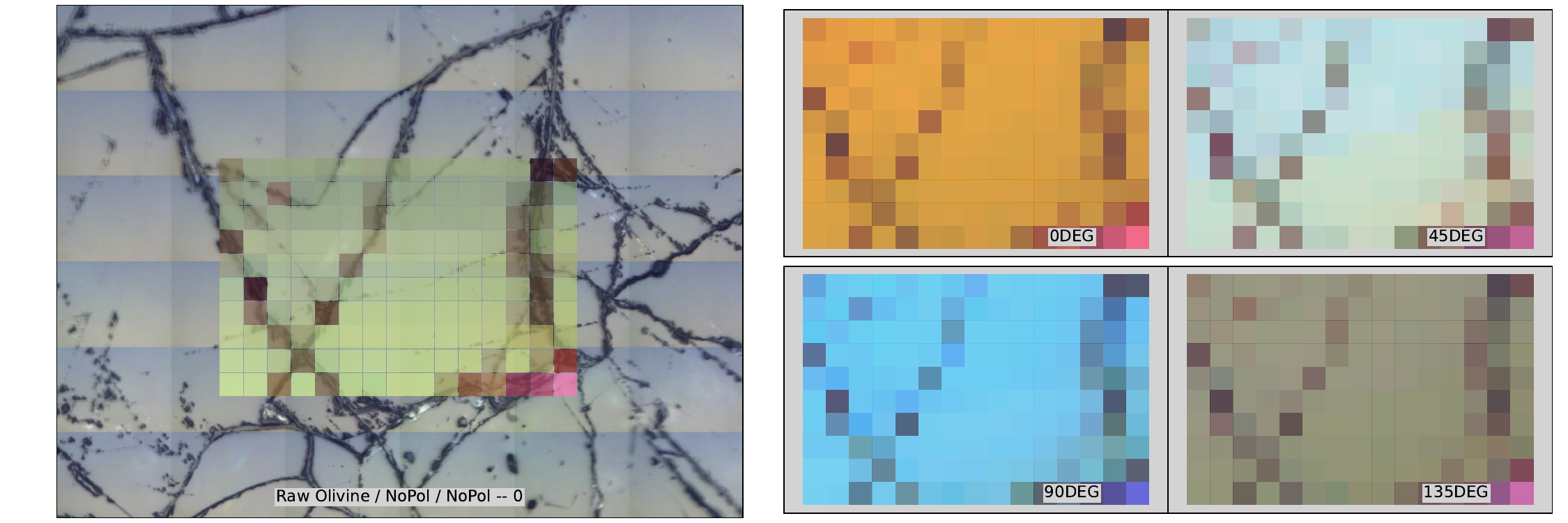}
  \caption{
  Selected spectrum is highlighted on the map by a rectangular profile. The optical image is collated from $250\times 200~\mu$m$^2$ $(x,y)$ micro-frames; the pixel of IR spectral acquisition was $\sim 40\times 40~\mu$m$^2$.
  Top: Synchrotron reflectance FTIR spectra obtained with non-polarised IR beam (left), and with polarised IR beam at polarisation angles of $0^o, 45^o, 90^o,$ and $135^o$ (right). Bottom: The optical microscopic image showing the total mapped area of $750\times 500~\mu$m$^2$ with each pixel representing the area of $50\times 50~\mu$m$^2$.}
  \label{f-Raw_Olivine}
\end{figure*}

\begin{figure*}[htbp]
  \centering
  \includegraphics[width=0.89\textwidth]{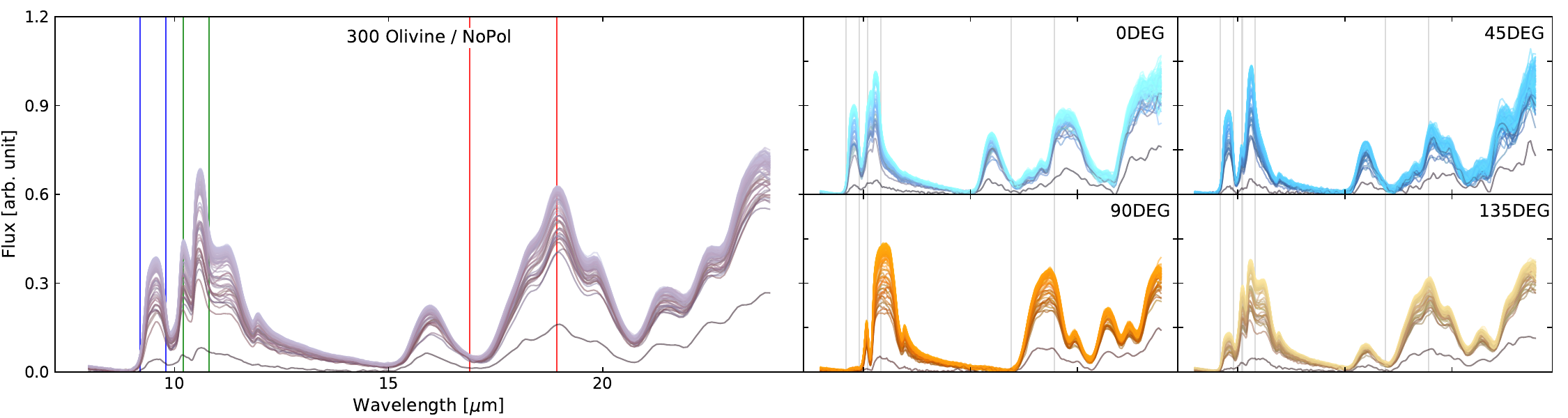}
  \hfill
  \includegraphics[width=0.89\textwidth]{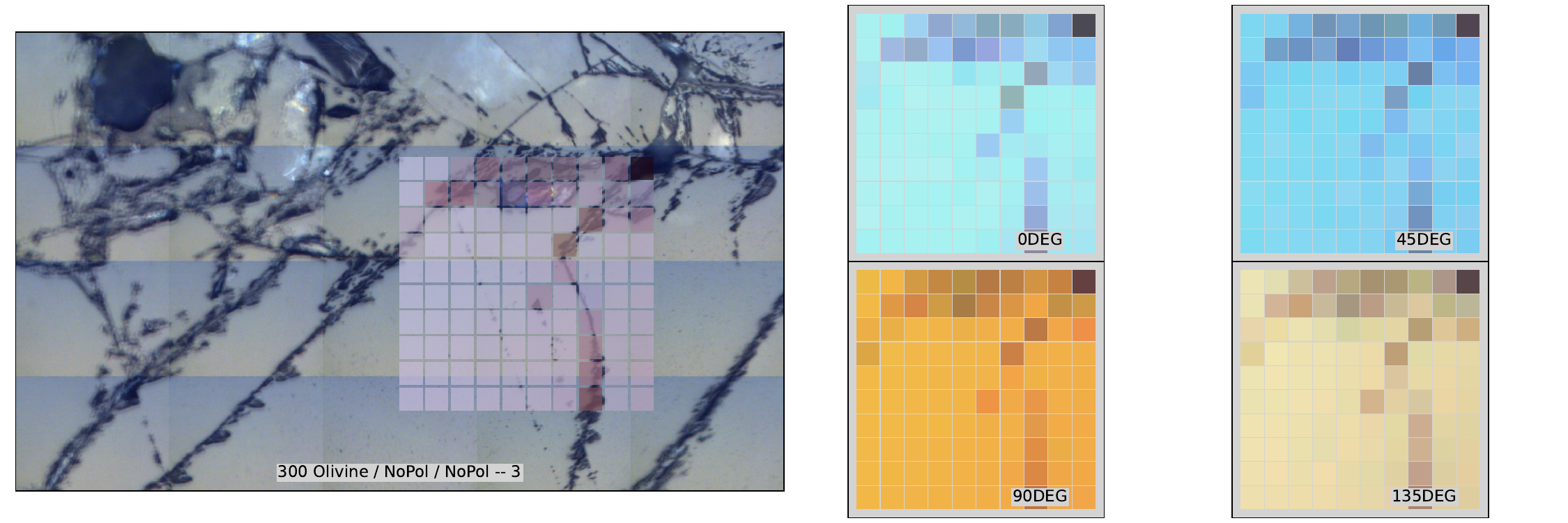}
  \caption{300$^\circ$C annealed Olivine: Spectra (left) and Map (right).}
  \label{f-300_Olivine}
\end{figure*}

\begin{figure*}[htbp]
  \centering
  \includegraphics[width=0.89\textwidth]{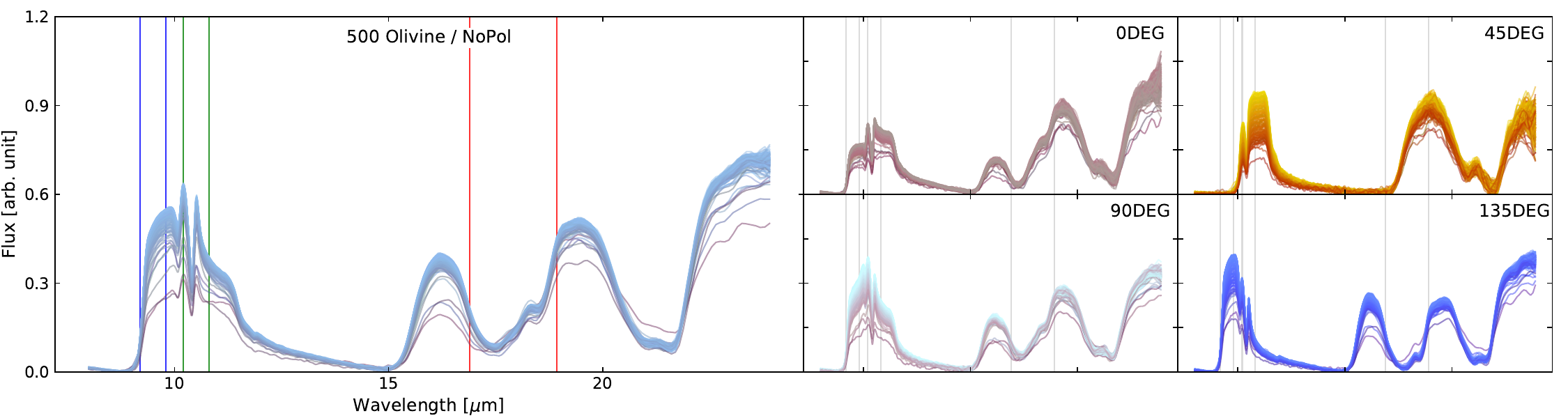}
  \hfill
  \includegraphics[width=0.89\textwidth]{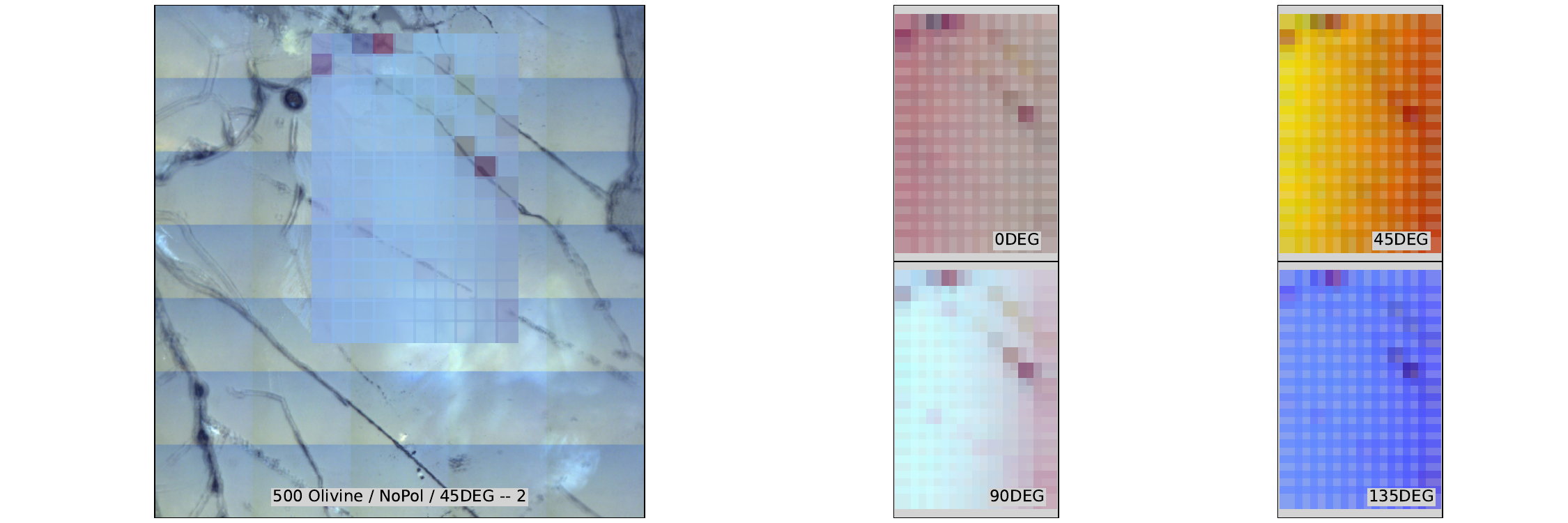}
  \caption{500$^\circ$C annealed Olivine: Spectra (left) and Map (right).}
  \label{f-500_Olivine}
\end{figure*}

\begin{figure*}[htbp]
  \centering
  \includegraphics[width=0.89\textwidth]{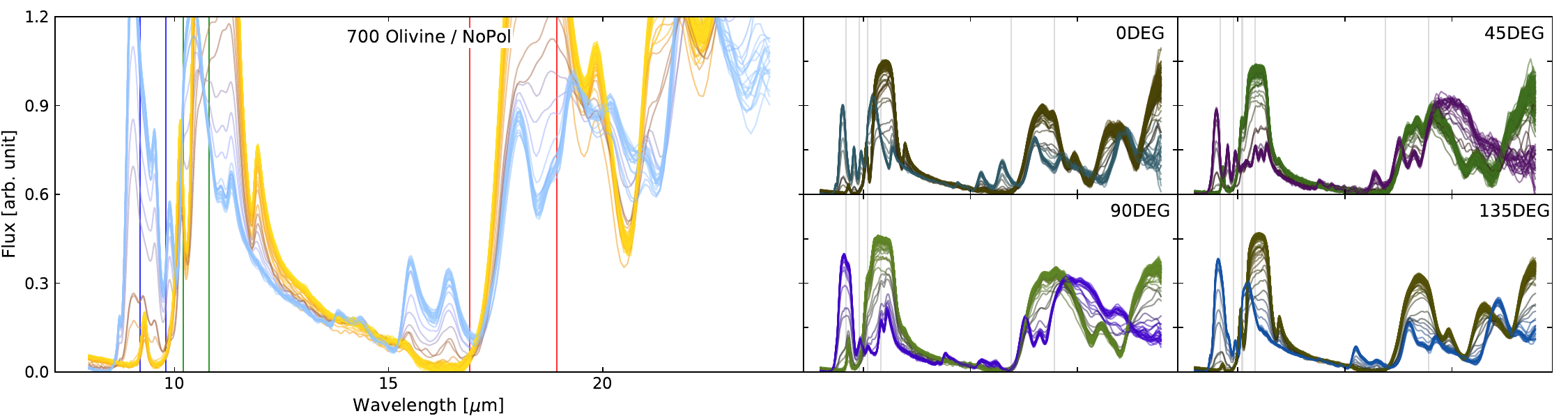}
  \hfill
  \includegraphics[width=0.89\textwidth]{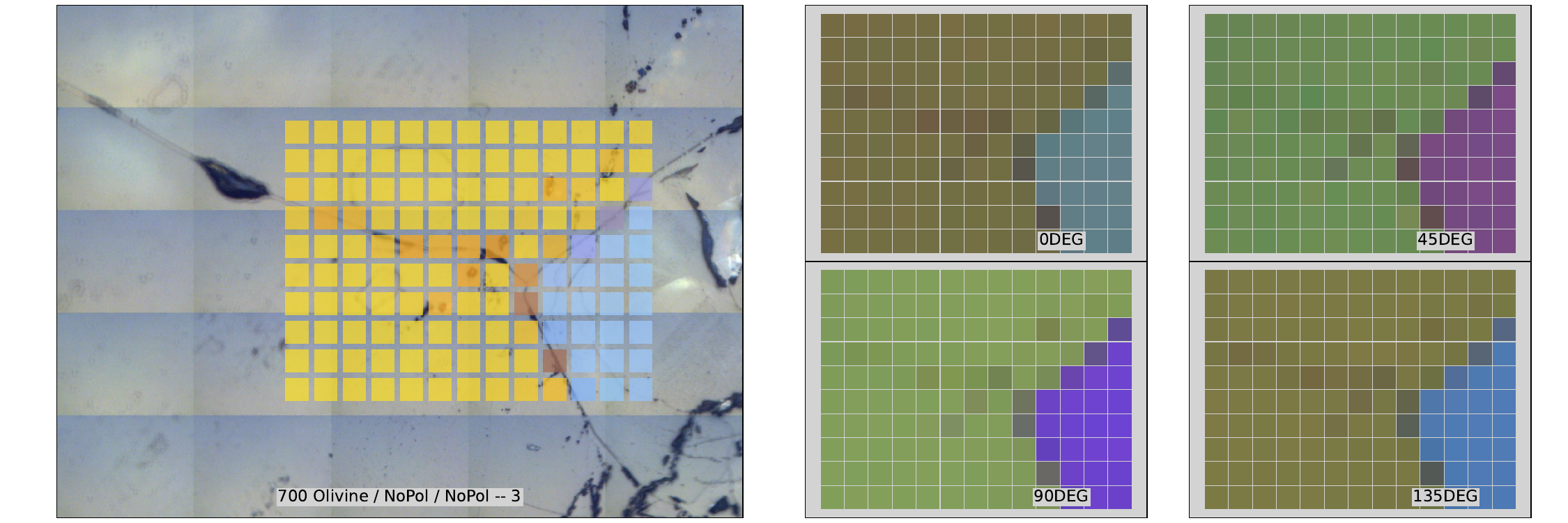}
  \caption{700$^\circ$C annealed Olivine: Spectra (left) and Map (right).}
  \label{f-700_Olivine}
\end{figure*}

\begin{figure*}[htbp]
  \centering
  \includegraphics[width=0.89\textwidth]{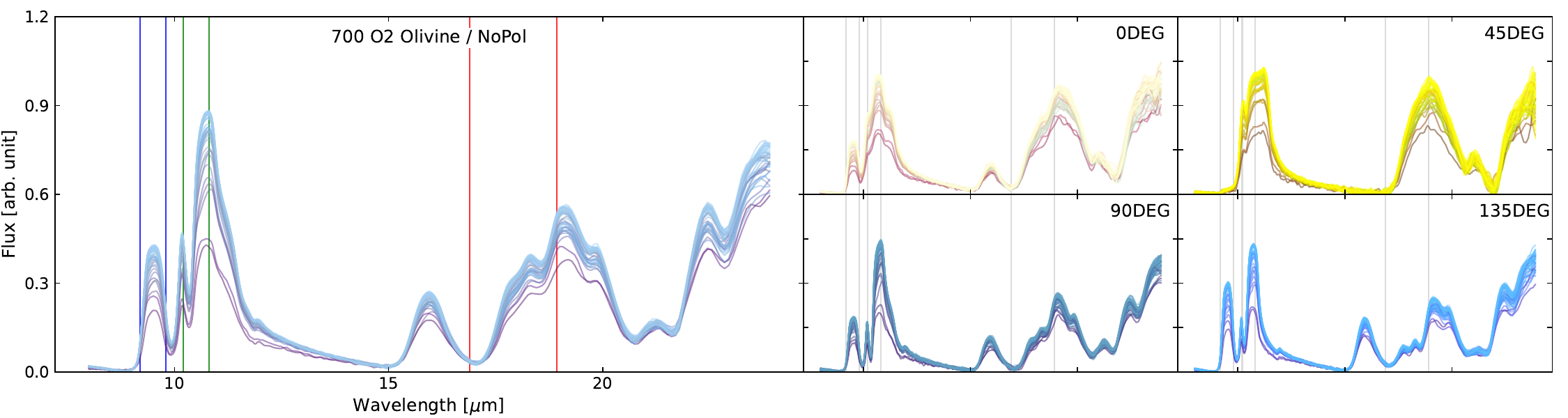}
  \hfill
  \includegraphics[width=0.89\textwidth]{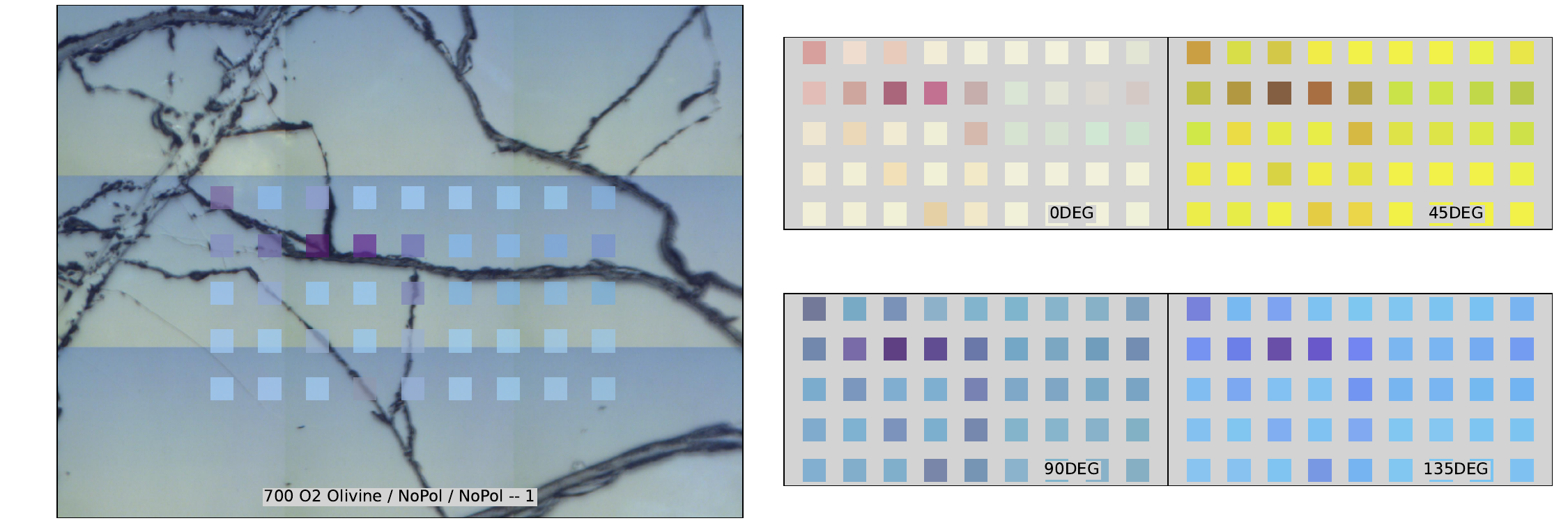}
  \caption{700$^\circ$C annealed \ce{O2} Olivine: Spectra (left) and Map (right).}
  \label{f-700_O2_Olivine}
\end{figure*}

\begin{figure*}[htbp]
  \centering
  \includegraphics[width=0.89\textwidth]{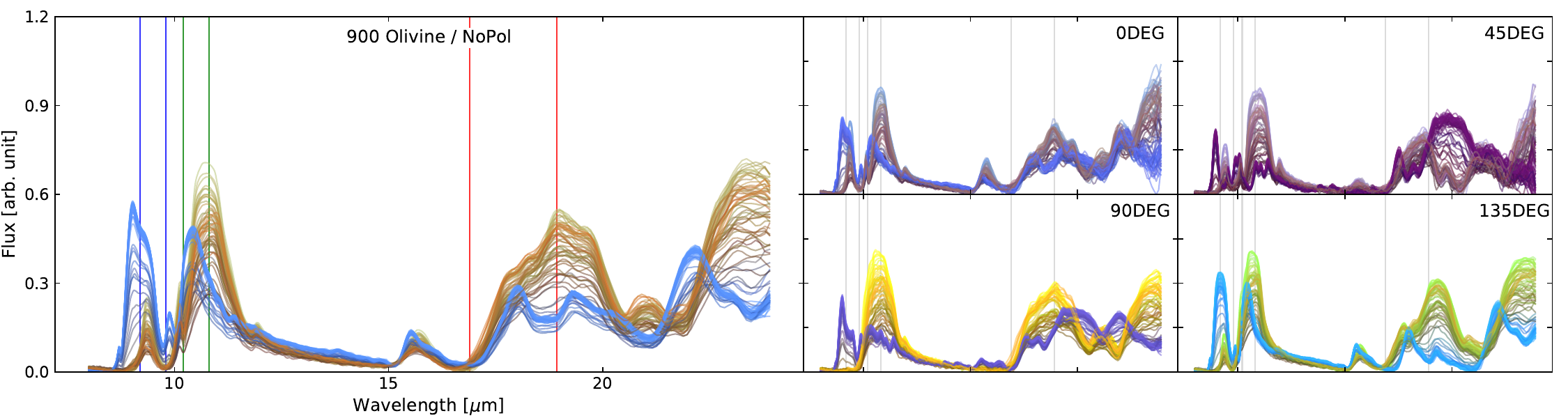}
  \hfill
  \includegraphics[width=0.89\textwidth]{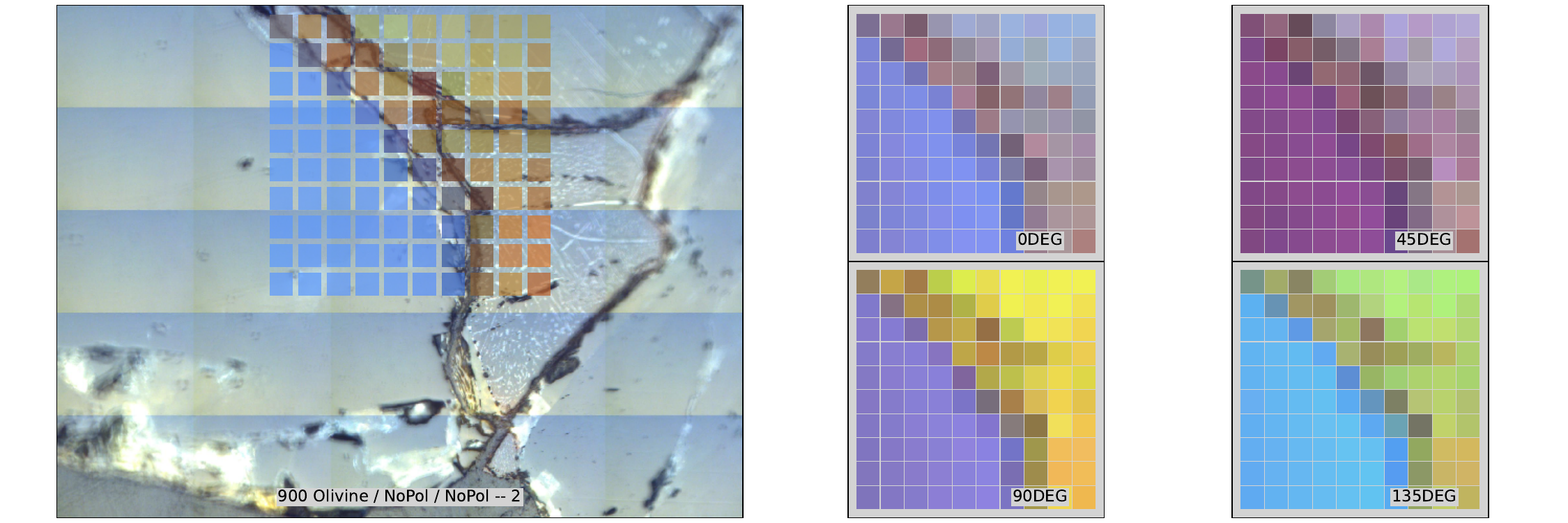}
  \caption{900$^\circ$C annealed Olivine: Spectra (left) and Map (right).}
  \label{f-900_Olivine}
\end{figure*}

\begin{figure*}[htbp]
  \centering
  \includegraphics[width=0.89\textwidth]{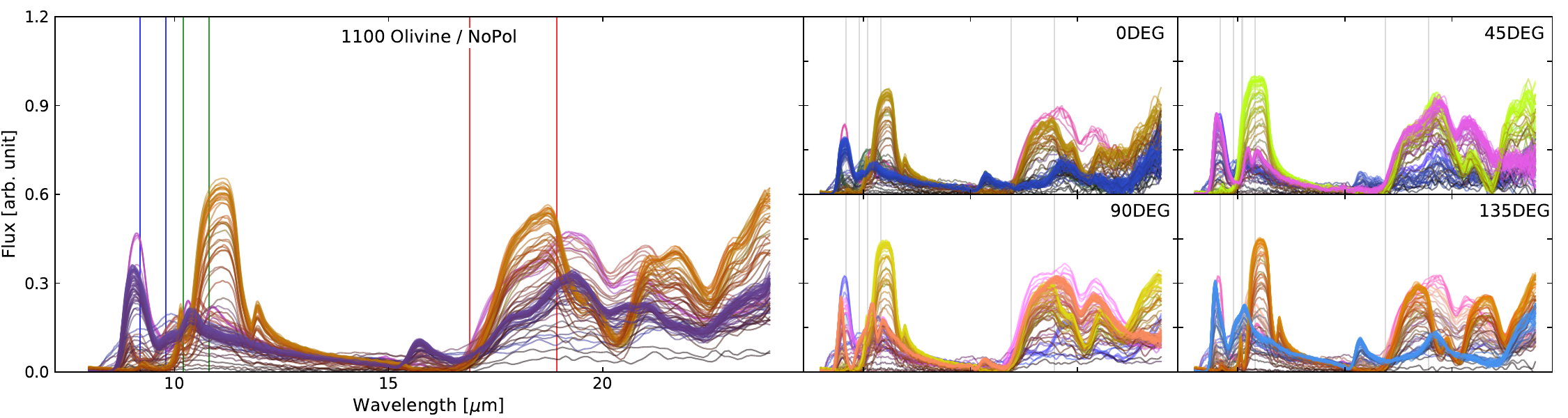}
  \hfill
  \includegraphics[width=0.89\textwidth]{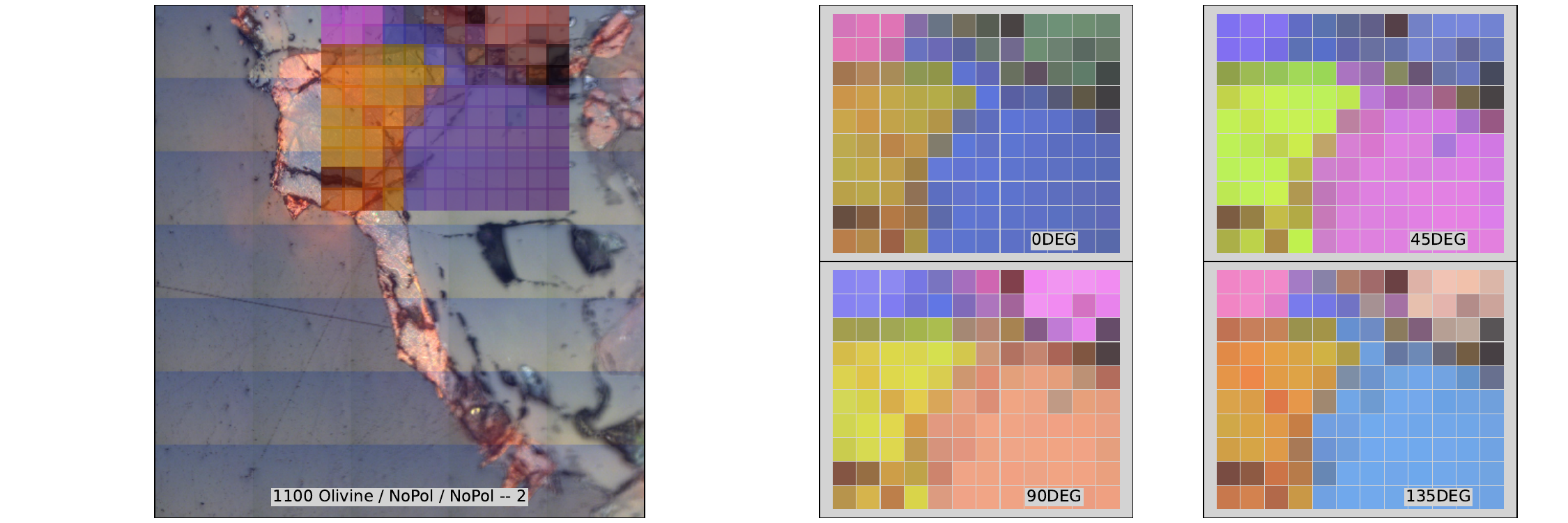}
  \caption{1100$^\circ$C annealed Olivine: Spectra (left) and Map (right).}
  \label{f-1100_Olivine}
\end{figure*}

\begin{figure*}[htbp]
  \centering
  \includegraphics[width=0.89\textwidth]{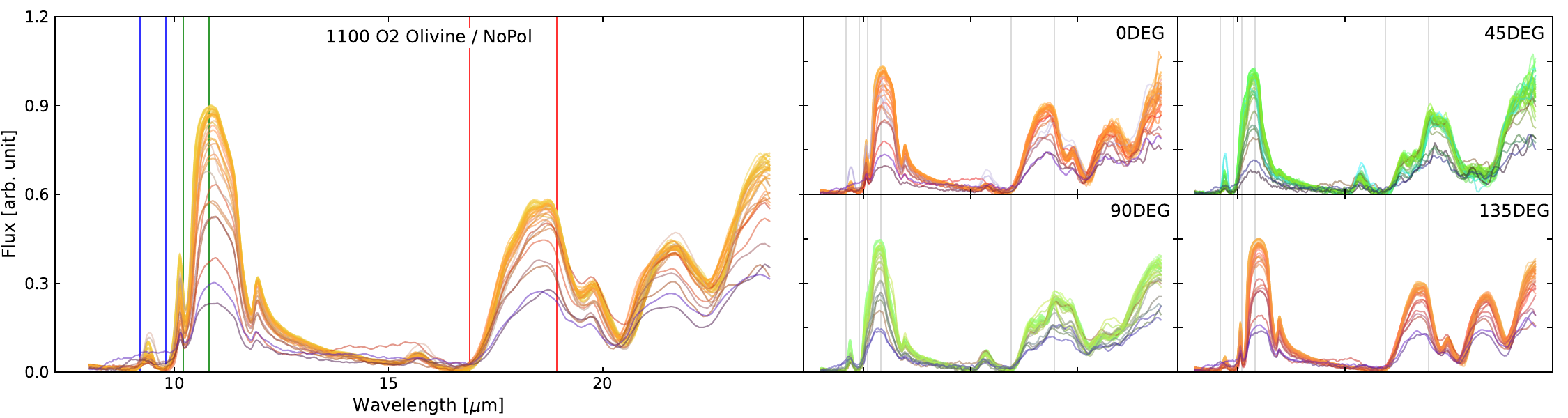}
  \hfill
  \includegraphics[width=0.89\textwidth]{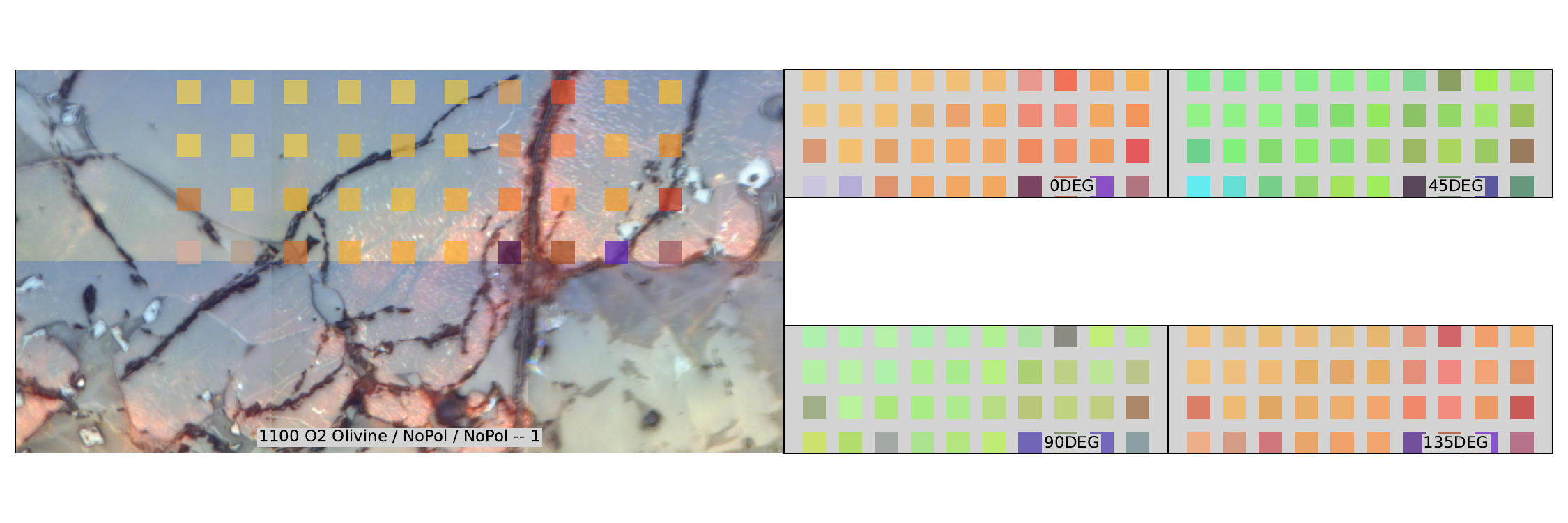}
  \caption{1100$^\circ$C annealed \ce{O2} Olivine: Spectra (left) and Map (right).}
  \label{f-1100_O2_Olivine}
\end{figure*}

\begin{figure*}[htbp]
  \centering
  \includegraphics[width=0.89\textwidth]{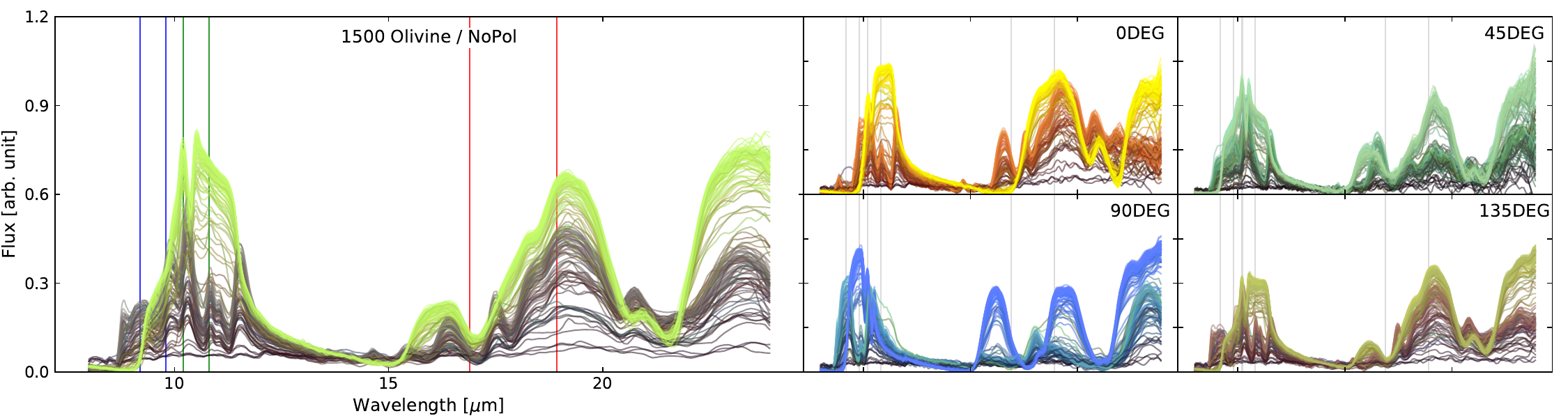}
  \hfill
  \includegraphics[width=0.89\textwidth]{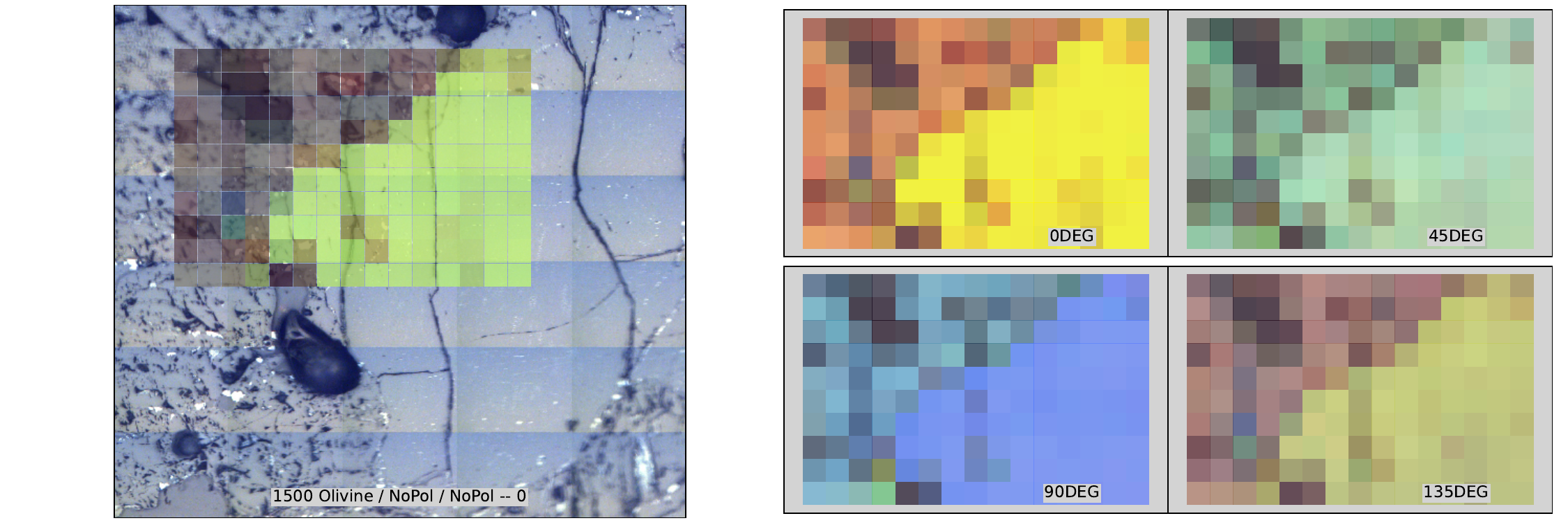}
  \caption{1500$^\circ$C annealed Olivine: Spectra (left) and Map (right).}
  \label{f-1500_Olivine}
\end{figure*}

\begin{figure*}[htbp]
  \centering
  \includegraphics[width=0.89\textwidth]{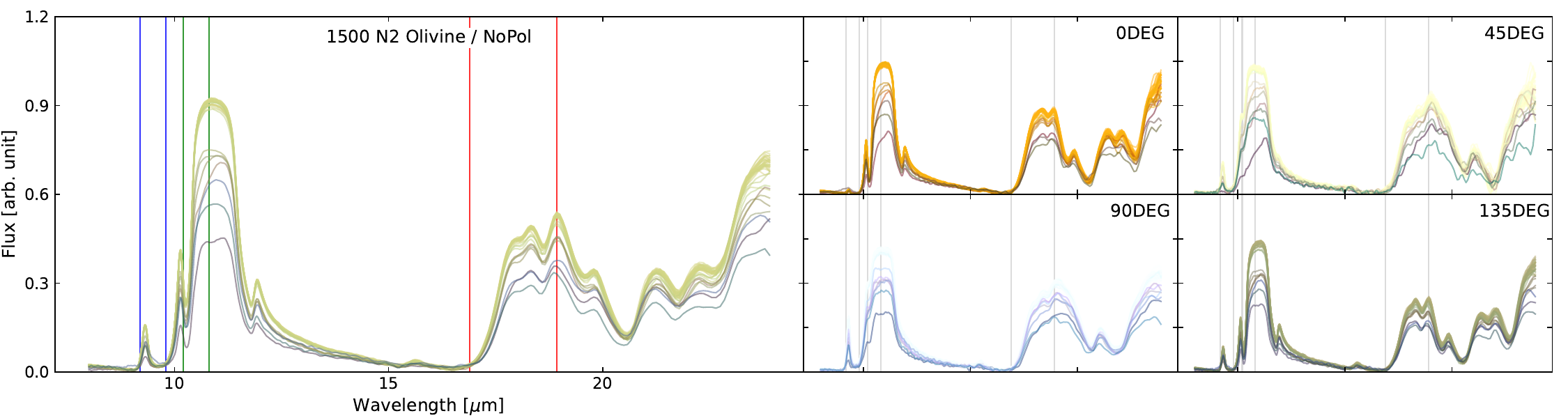}
  \hfill
  \includegraphics[width=0.89\textwidth]{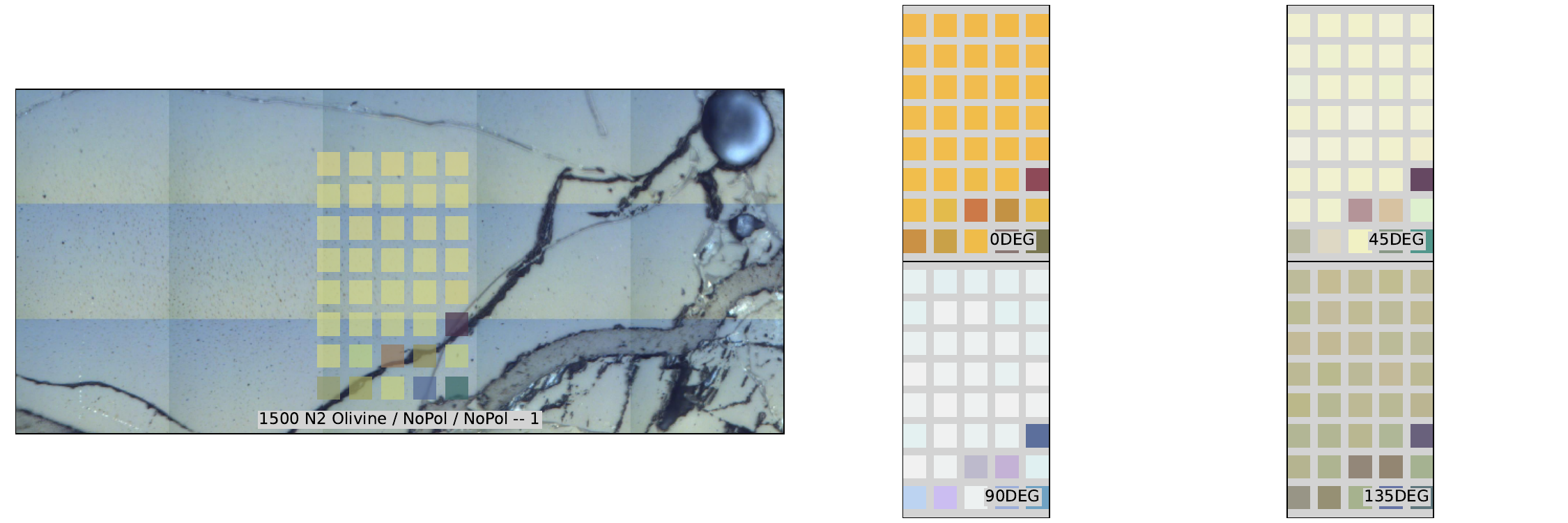}
  \caption{1500$^\circ$C annealed \ce{N2} Olivine: Spectra (left) and Map (right).}
  \label{f-1500_N2_Olivine}
\end{figure*}

\begin{figure*}[htbp]
  \centering
  \includegraphics[width=0.89\textwidth]{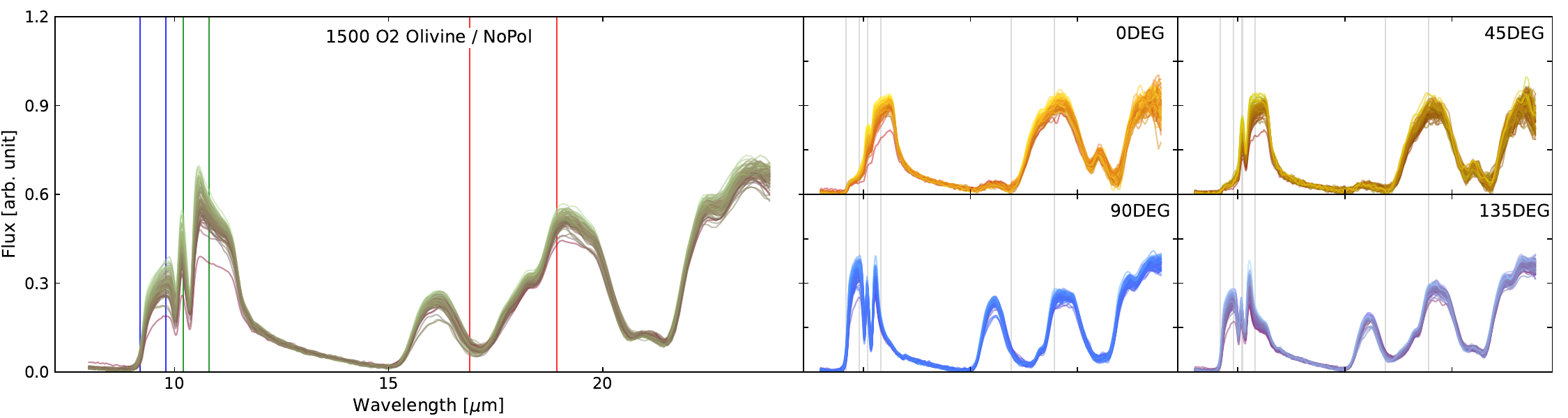}
  \hfill
  \includegraphics[width=0.89\textwidth]{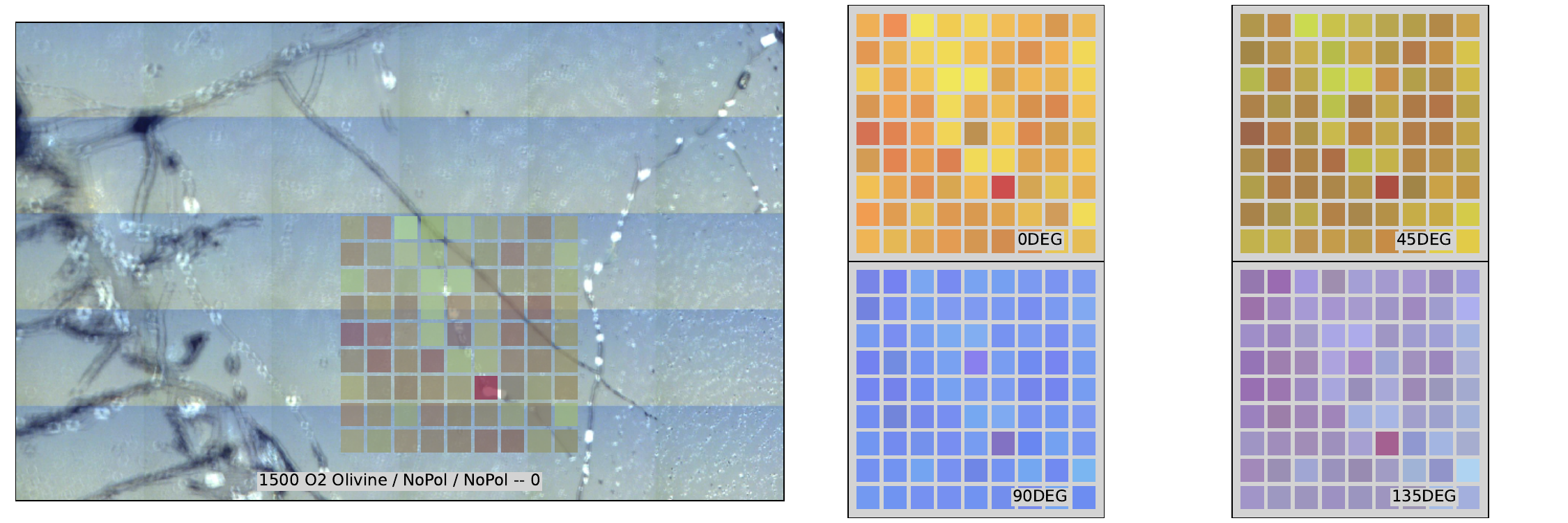}
  \caption{1500$^\circ$C annealed \ce{O2} Olivine: Spectra (left) and Map (right).}
  \label{f-1500_O2_Olivine}
\end{figure*}
\end{document}